\documentclass[a4paper]{scrbook} 
\usepackage{etex}
\usepackage{a4wide}
\usepackage{amsmath,amssymb,amsfonts,amsthm,makeidx,graphicx,xcolor}
\usepackage{txfonts}
\usepackage{helvet}
\usepackage{bm}
\usepackage[numbers]{natbib}
\usepackage[format=plain]{caption}
\usepackage[pdftex,pdftitle={Modeling Approaches and Computational
  Methods for Particle-laden Turbulent Flows (book chapter)},
breaklinks=true,bookmarks=true,colorlinks=true,linkcolor=blue,
citecolor=blue,allcolors=blue]{hyperref}
\usepackage{minitoc}
\usepackage[symbol]{footmisc}
\newif\ifnewauthor
\newauthortrue
\newcounter{affils}
\makeatletter
\def\chapterauthor[#1]#2{\ifnewauthor\else, \fi
  {\bfseries #2}\textsuperscript{\hyperref[authaffil\the\value{chapter}.#1]{#1}}%
  \ifnewauthor\newauthorfalse\gdef\chapterauthors{#2}\else
    \g@addto@macro\chapterauthors{, #2}\fi
  \ignorespaces 
}
\makeatother
\def\chapteraffil[#1]{%
\item[$^{#1}$]%
  \refstepcounter{affils}%
  \label{authaffil\the\value{chapter}.#1}%
}
\newenvironment{affils}
{\addtocontents{toc}{\chapterauthors\string\par}\begin{enumerate}}
  {\end{enumerate}\global\newauthortrue}
\newenvironment{abstract}{\rightskip1in 
{\bfseries Abstract:} }{}
\usepackage{scrlayer-scrpage}
\automark{chapter}
\lehead{
  Modelling approaches and computational methods for particle-laden turbulent flows}
\rohead{\leftmark}
{\rightskip1in 
{\bfseries Acknowledgements:} }{}
\usepackage{ulem}
\usepackage{upgreek}

\def\solid{\protect\rule[1pt]{10.pt}{1pt}}

\def\solidshort{\protect\rule[2pt]{3.pt}{1pt}}
\def\dashed{\solidshort$\,$\solidshort$\,$\solidshort}

\definecolor{darkgreen}{RGB}{0,102,0}
\newcommand{\revision}[2]{#2}
\newcommand{\revisions}[1]{}
\newcommand{\proofedit}[2]{#2}
\newcommand{\proofedits}[1]{}
\usepackage[percent]{overpic}
\usepackage{relsize}
\definecolor{ACgreen}{rgb}{0.1, 0.7, 0.1}
\definecolor{ACdarkgreen}{rgb}{0.05, 0.35, 0.05}
\definecolor{ACdarkred}{rgb}{0.6, 0.0, 0.0}
\definecolor{ACyellow}{rgb}{0.9 0.7 0.2}
\definecolor{ACpurple}{rgb}{0.7 0.1 0.9}
\definecolor{ACcyan}{rgb}{0.0 0.75 0.75}
\definecolor{ACblue}{rgb}{0.0 0.5 0.9}
\definecolor{ACdarkblue}{rgb}{0.0 0.25 0.45}
\definecolor{ACred}{rgb}{0.85 0.0 0.0}
\newcommand{\ACcirc}{$\boldsymbol{\circ}$}
\newcommand{\ACcircfull}{$\bullet$}
\newcommand{\ACsquare}{$\boldsymbol{\square}$}
\newcommand{\ACtriup}{$\boldsymbol{\triangle}$}

\newcommand{\ACdiamond}{$\boldsymbol{\diamond}$}
\newcommand{\ACtrileft}{$\triangleleft$}
\newcommand{\ACtridown}{$\boldsymbol{\triangledown}$}
\newcommand{\ACcross}{$\boldsymbol{\times}$}
\usepackage{tikz}

\newcommand{\bu} {{\mathbf{u}}}

\newcommand{\bee}{\begin{equation}}
\newcommand{\eee}{\end{equation}}
\newcommand{\bea}{\begin{eqnarray}}
\newcommand{\eea}{\end{eqnarray}}

\newcommand{\lab}{\langle}
\newcommand{\rab}{\rangle}

\begin{document}
\epstopdfsetup{suffix=} 
%
\setcounter{chapter}{5} %
  \chapter[Results obtained with PR-DNS]
  {%
    Results from Particle-Resolved Simulations\footnote{%
      Published as Chapter~6 in {\itshape
        Modeling Approaches and Computational Methods for
        Particle-laden Turbulent Flows 
      },
      S.\ Subramaniam and S.\ Balachandar (editors), Academic press, 
      2022, 
      \url{https://doi.org/10.1016/B978-0-32-390133-8.00013-X}.
    }
  }
  \label{PRR_chap6} 
\chapterauthor[1]{Agathe Chouippe}%
\chapterauthor[2]{Aman G. Kidanemariam}%
\chapterauthor[3]{Jos Derksen %
}%
\chapterauthor[4,5]{Anthony Wachs}%
\chapterauthor[6]{Markus Uhlmann}
  \begin{affils}
\chapteraffil[1]{%
  {Universit\'e de Strasbourg},
  {Institut ICube, Fluid Mechanics Group},
  {2, rue Boussingault, 67000 Strasbourg, France}} 
\chapteraffil[2]{%
  {The University of Melbourne}, 
  {Department of Mechanical Engineering}, 
  {Victoria 3010, Australia}}
\chapteraffil[3]{%
    {University of Aberdeen},
  {School of Engineering},
  {King's College, Aberdeen AB24 3UE, United Kingdom}}
\chapteraffil[4]{%
    {University of British Columbia},
  {Department of Mathematics},
  {1984 Mathematics Road, Vancouver BC V6T 1Z2, CANADA}}
\chapteraffil[5]{%
    {University of British Columbia},
    {Department of Chemical and Biological Engineering},
    {2360 East Mall, Vancouver BC V6T 1Z3, Canada}
  }  
  \chapteraffil[6]{%
    {Karlsruhe Institute of Technology},
  {Institute for Hydromechanics},
  {Kaiserstr. 12, 76131 Karlsruhe, Germany}}
%
%
%
  \end{affils}
%
%
%
%
%
\minitoc
\begin{abstract}
  We review some of the results obtained to date with the aid of the
  PR-DNS approach to turbulent particulate flows.
  It is shown that the method has matured to a point which allows to
  apply it successfully to a wide variety of fluid/particle
  configurations, albeit still at a relatively large computational
  cost. 
  Due to the availability of high-fidelity space-and-time-resolved
  data, a number of challenging open questions have already been
  addressed in unprecedented detail. 
\end{abstract}

%
%
%
%

%
%
%
%
%
%
%
%
%
%
\section{Introduction}
\label{sec:PRR-intro}
Since its
emergence 
in the late 1990s the method of
particle-resolved direct numerical simulation (PR-DNS) 
has established itself as an eminently useful source of high-fidelity
data on turbulent particulate flow.
In the PR-DNS approach the Navier-Stokes equations for the fluid flow
are solved together with the Newton-Euler equations for rigid body
motion (cf.\ chapter~5) 
without further modelling
assumptions -- except for those which relate to solid-solid contacts,
cf.\ chapter~7. 
Results obtained with this method
thereby provide insight into the micro-scale flow around each
particle, and on the way these scales couple back to the remainder of
the spectrum.  

Despite the formidable computational requirements inherent in the
PR-DNS approach, a growing body of literature involving such numerical
studies has been published over the last two decades.
The results have already had a significant impact on the state of the
art in particulate flow: (i) they have helped to clarify the
mechanisms behind various physical phenomena,
and (ii) they have aided in the design and validation of simplified
models (e.g.\ of Euler-Lagrange or Euler-Euler type).
It is the purpose of the present chapter to provide an overview of the
capabilities and current limitations of the PR-DNS approach. Many of
the topics touched upon in the following will be picked up in
subsequent chapters of the present volume, such as
chapter~9 
on extended point particle models and
chapter~14 
on the modelling of fluidized beds.

Since the relevant parameter space is large
(\revision{}{particle} volume fraction, Galileo number, density ratio, length scale
ratio)  
and 
particulate flow applications are highly diverse 
(laminar vs.\ turbulent background flow, homogeneous vs.\
inhomogeneous systems, wall-bounded vs.\ unbounded, fixed vs.\ mobile
particles, turbulence forcing method),
the following review can only cover a small subset of this vast
topic. 
Here we have chosen to focus on three aspects.
First, \revision{}{in section~\ref{sec:PRR-drag-models}}
we will give an account of how PR-DNS can be used for the
purpose of improving models for quasi-steady drag and for the
analogous scalar transfer. 
Section~\ref{sec:PRR-unbounded} is dedicated to fluid-mediated
particle dynamics in
\revision{}{dilute}
unbounded flow with and without a priori
turbulence.
Finally, we turn our attention to wall-bounded shear flow in 
section~\ref{sec:PRR-wall-bounded}, where we discuss vertical and
horizontal channel flow, the latter one in the absence or presence of
a thick sediment bed, possibly featuring macroscopic patterns. 
\section{PR-DNS \revision{}{of dense fluidized systems} for drag force parameterizations  based on dynamic 
  simulations}
\label{sec:PRR-drag-models}
Particle-resolved simulations of suspensions in \revision{}{triple periodic} domains have been performed to probe solids-fluid
and solids-solids momentum transfer and \textendash{} more generally
\textendash{} the evolution of the microstructure of suspensions. In
this section, we will restrict ourselves to systems \revision{}{comprised} of a
Newtonian fluid and solid particles all having the same size. This
restriction is for brevity mostly. Simulations with spherical
particles having a size distribution as well as simulations with
non-Newtonian fluids have been reported
\cite{vanderhoef2005,derksen2009}.
The essentials of their setup and interpretation can be understood in
the context of what is described in this section. Particle-resolved
simulations involving non-spherical particles are a strongly emerging
field of study and
Chapter~7 
is largely dedicated to simulation
methodologies for this subject. At the end of this section we will
briefly return to the topic of non-spherical particles. 

A number of distinctions for the already quite specific flow
configurations still need to be made.  

In the simplest configuration, the particles are kept at fixed
locations and fluid flows through the particle assemblies under the
influence of a constant and uniform body force. Such simulations date
back to
\citet{ladd1994b}.
Their most relevant application
is in probing fluid-particle drag on spheres as a function of the
\revision{}{particle} volume fraction, the topology of the particle arrangement
(ordered versus random assemblies), and a particle-based Reynolds
number. The availability of analytical results for drag on various
types of regular static assemblies of spheres at low Reynolds numbers
\cite{sangani1982,hasimoto1959} greatly facilitates the verification
of the numerical approaches.  

Drag in static particle assemblies has practical implications in
gas-solid flows, i.e. flows with high Stokes numbers. Then, the time
scales over which the particle configurations evolve are much larger
than fluid flow time scales at the particle-level (e.g. particle size
over superficial velocity) so that a frozen-particle approach is
justified. At lower Stokes numbers (e.g. liquid-solid flows) particle
configurations respond more directly to flow conditions which
\textendash{} as we will see \textendash{} has consequences for the
drag force. 

 For specific numerical methodologies as well as results of flow
 through static particle assemblies \textendash{} including drag force
 correlations derived from simulations \textendash{} we refer to the
 chapter on fluidization (Chapter~14) 
 in this book.  

In this section the focus is on simulations of freely moving,
colliding and rotating spheres in fluid and the information that can
be extracted from such simulations. In these systems, agitation can be
done in different ways. \textbf{(1)} As a direct and logical extension
of flows through static solids assemblies, a uniform and constant body
force is applied to the fluid. Since the fully periodic system needs
to be overall force-balanced, an opposing force is applied to the
particles. This usually is interpreted as a fluidized system with the
particles feeling gravity and the interstitial fluid pushed through by
the body force in the opposite direction
\cite{derksen2007}. \textbf{(2)} By creating turbulence in fully 
periodic domains, one is able to directly study particle-turbulence
interactions. Creating single-phase turbulence can be done through
forcing \cite{alvelius1999}. Placing particles in fully developed
single-phase turbulence and letting the system decay enabled Lucci et
al \cite{lucci2010} to highlight finite particle size effects
\textendash{} specifically for spheres with a diameter larger than the
Kolmogorov scale \textendash{} on the evolution of turbulent kinetic
energy (see Chapter~3). 
Dynamically steady turbulent solid-liquid
systems \cite{tencate2004} with resolved particles show turbulence
modulation consisting of the transfer of turbulent kinetic energy from
large scales to length scales of the order of the particle size. 

In this section we will provide more detail and results regarding
fluidized dynamic systems with solid over fluid density ratios
characteristic of liquid fluidization. Full numerical details can be
found in \cite{derksen2007}. The fluid has density $\rho_{f}$ and
kinematic viscosity $\nu $; the particles are solid spheres all having
diameter $d_{p}$ and density $\rho_{p}$. The \revision{}{particle} volume fraction is
$\phi$. This allows to define a mixture density as ${\rho_{m}
}=\phi \rho _{p}+\left(1-\phi \right)\rho_{f} $. Gravitational
acceleration $\mathbf{g}$ points in the negative \textit{z}-direction
of a Cartesian coordinate system:
$\mathbf{g}=-g\mathbf{e}_{\mathbf{z}}$.
The above definitions give
rise to a set of three dimensionless input parameters governing the
system. Here we select $\phi $, density ratio $\rho _{p}/\rho_{f}$ and a
Galileo number ${Ga}=((\rho_{p}/\rho_{f}-1)d_{p}g)^{1/2}d_{p}/\nu$ as the input parameters.  

Each particle experiences a net gravity force
$\mathbf{F}_{\mathbf{g}}=-\left(\rho _{p}-\rho_{m}
  \right)\frac{\pi }{6}d_{p}^{3}g\mathbf{e}_{\mathbf{z}}$. An overall
force balance then requires a body force acting on the fluid
$\mathbf{f}=\left({\rho_{m} }-\rho_{f}
\right)g\mathbf{e}_{\mathbf{z}}$. In addition to gravity, the
particles feel hydrodynamic forces and forces due to close-range
particle-particle interactions: radial lubrication and a spring-type
force to deal with solid-solid contact (``soft'' collisions, 
Chapter~7 
in this book).  

As an alternative for soft collisions, a hard-sphere collision
algorithm can be implemented \cite{derksen2007}. This has the
advantage of particles not overlapping so that overall \revision{}{particle} volume
fraction levels are not being compromised. For dense suspensions
($\phi \geq 0.4$) and/or systems containing many particles ($N\geq
10^{4}$), hard-sphere collision algorithms slow down the computations
significantly given that each collision needs to be accounted for
explicitly and that parallelization of the algorithm requires
significant communication. No strong sensitivity of simulation results
with respect to collision parameters (restitution and friction
coefficients) has been observed for solid-liquid flow
\cite{derksen2007}. This is because most dissipation between
approaching particles takes place in the liquid prior to collision,
either in the resolved flow or as a result of lubrication modelling.  

Simulations are initialized by randomly placing the equally sized
spheres in the fully periodic, three-dimensional domain in a
non-overlapping manner. There is a limit to the \revision{}{particle} volume fraction
that can be achieved this way: $\phi \leq 0.3$
\cite{torquato2000}. Higher \revision{}{particle} volume fractions can be achieved by
a compaction process: first build a random particle configuration with
$\phi <0.3$ and then let the particles freely move and collide under
the influence of a force that attracts them to a centre plane of the
domain. Once a desired volume fraction has been achieved, the parts of
the volume void of particles are discarded. Alternatively, high \revision{}{particle}
volume fractions can be generated by ``growing'' the particles: create
an initial non-overlapping random set of spheres with $\phi <0.3$;
give them random velocities and let them move and collide as a
granular gas. Each time step \revision{}{one} determines the minimum distance $\delta
_{min}$ between two particle surfaces in the entire domain and
\revision{}{then} increases the radius of each sphere by an amount slightly smaller than
$\delta _{min}/2$. This way \revision{}{particle} volume fractions up to those
associated to random close packing ($\phi \approx 0.62$) can be
accomplished. 

\begin{figure}
  \centering
  \includegraphics[width=.4\linewidth]
  {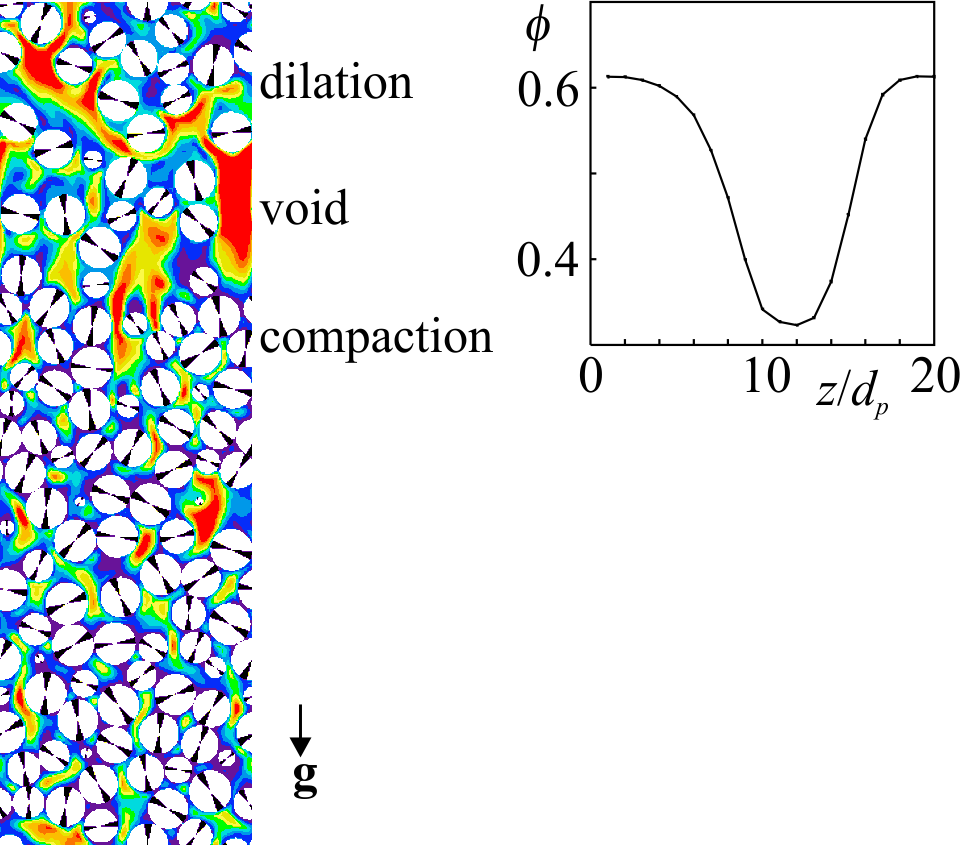} 
  \caption{
    Wave instability in liquid fluidized bed
    \cite{derksen2007}. Overall \revision{}{particle} volume fraction $\langle
      \phi\rangle =0.51$, $\rho _{p}/\rho_{f}=4.4$ and
    ${Ga}=133$. Left: cross
    section of an instantaneous realization with liquid coloured by
    velocity magnitude; right: average wave form.} 
  \label{fig:PRR-waves}
\end{figure}

After initializing the two-phase system, the forces on fluid and
solids are activated and we let the system evolve to a dynamically
steady state. Under certain conditions and \textendash{} most
importantly \textendash{} with a sufficiently large domain, wave and
void-type instabilities develop \cite{duru2002}. In sufficiently
narrow domains, the characteristics of one-dimensional waves (their
speed \textit{c} and wave form) have been used as a way of validating
simulation results with available experimental data
\cite{derksen2007}. Figure~\ref{fig:PRR-waves}  shows an impression of a
simulation after having developed a planar wave. The dimensionless
wave speed $\tilde{c}=cd_{p}/\nu $ in experiment \cite{duru2002} and
simulation \cite{derksen2007} under the conditions (in dimensionless
terms) as given in the caption of Figure~\ref{fig:PRR-waves} are
$\tilde{c}=29\pm 1$ and $\tilde{c}=33\pm 2$ respectively. 

Interpretation of simulations in terms of average drag requires
determining the average interstitial fluid and solids speed (averaging
over volume and time), $\left\langle \mathbf{u}\right\rangle $ and
$\left\langle \mathbf{u}_{\mathbf{p}}\right\rangle $ respectively. The
superficial slip velocity then is $\left(1-\phi \right)\left|
  \left\langle \mathbf{u}\right\rangle -\left\langle
    \mathbf{u}_{\mathbf{p}}\right\rangle \right| $ which defines a
Reynolds number as ${Re}={\left(1-\phi \right)\left|
    \left\langle \mathbf{u}\right\rangle -\left\langle
      \mathbf{u}_{\mathbf{p}}\right\rangle \right| d_{p}}/{\nu }$. With a
constant net gravity force acting on each particle and \textendash{}
on average \textendash{} balanced forces
on each particle the hydrodynamic force per particle is
$-\mathbf{F}_{\mathbf{g}}$. This then allows for the definition of a
dimensionless hydrodynamic force coefficient as $F=\frac{\left(\rho
    _{p}-{\rho_{m} }\right)\frac{\pi }{6}d_{p}^{3}g}{3\pi d_{p}\nu \rho_{f}
  \left(1-\phi \right)\left| \left\langle \mathbf{u}\right\rangle
    -\left\langle \mathbf{u}_{\mathbf{p}}\right\rangle \right| }$
which is the drag force normalized by Stokes drag based on the
superficial slip speed. 

From a physics perspective, the average dimensionless drag in a
homogeneous system ($F$) depends on ${Re}$, $\phi $ and the
density ratio $\rho _{p}/\rho_{f}$. Instead of the density ratio, the
Stokes number is regularly used as an independent dimensionless
variable: ${St}=\frac{1}{18}\frac{\rho _{p}}{\rho_{f}
}{Re}$, so that $F\left({Re,}\phi
  ,{St}\right)$. From a numerical perspective one also needs to
\revision{}{consider} effects of domain size, spatial (grid) resolution
and time step.  

As noted above, there is extensive literature on
$F\left({Re,}\phi ,{St}\rightarrow \infty \right)$ based
on PR-DNS’s of flow through fixed particle assemblies. The relevance of
the Stokes number has been identified in \cite{rubinstein2016} where
it is noted that over the full range of \revision{}{particle} volume fractions of
random assemblies ($0<\phi \leq 0.62$), and in the low Reynolds number
limit, the dimensionless drag force according to fixed-particle
simulations by Van der Hoef et al \cite{vanderhoef2005}
$F=\frac{10\phi }{1-\phi }+\left(1-\phi \right)^{3}\left(1+1.5\phi
  ^{1/2}\right)$ is higher than the empirical correlation by Wen and
Yu \cite{wen1966} $F=\left(1-\phi \right)^{-2.65}$. The latter
correlation is based on liquid fluidization experiments at moderate
Stokes numbers. The moderate solids inertia allows the particles to
quickly adapt their linear and angular velocity and location such that
flow resistance gets reduced as confirmed by Rubinstein et al
\cite{rubinstein2016}. By performing PR-DNS of freely moving spherical
particles in periodic domains over a range of density ratios, they
quantified \textendash{} by fitting average drag results of dynamic
PR-DNS’s \textendash{} the way the transition between high and low Stokes
number occurs:  
\begin{equation}
F=\alpha \left(\widetilde{{St}}\right)F_{hi{St}}+\left[1-\alpha \left(\widetilde{{St}}\right)\right]F_{WY},
\label{equ:PRR-f_stokes}
\end{equation}
with $F_{hi{St}}$ the high-Stokes limit for which
\cite{rubinstein2016} took the Van der Hoef et al
\cite{vanderhoef2005} correlation, $F_{WY}$ the Wen and Yu
\cite{wen1966} correlation and  

\begin{equation}
\alpha \left(\widetilde{{St}}\right)=\frac{1}{2}\left(1+\frac{\widetilde{{St}}-10}{\widetilde{{St}}+10}\right),
\label{equ:PRR-alfa}
\end{equation}
with $\widetilde{{St}}=\frac{{St}}{\left(1-\phi
  \right)^{2}}$. An extension towards moderate Reynolds numbers (up to
100) of the effect of solids inertia through the density ratio on
particle drag has been presented in a comprehensive study by
Tavanashad et al \cite{tavanashad2021} based on particle resolved
simulations. Drag force correlations play an important role in
Euler-Lagrange simulations, see Chapter~13. 

Clustering and preferential relative particle locations at moderate
Stokes numbers not only have consequences for the drag force, it also
impacts mass transfer. This is relevant with a view to the application
of suspensions in (chemical) engineering processes to achieve mass
transfer between a fluid and a solids phase.  

This has been investigated in \cite{derksen2014} where \textendash{}
in the same way as discussed above \textendash{} the joint dynamics of
fluid and solids in fully periodic domains driven by a uniform body
force has been investigated. Once the solid-fluid system is fully
developed one starts solving a convection diffusion equation  
\begin{equation}
\frac{\partial c}{\partial t}+\mathbf{u}\cdot \nabla c=\Gamma \nabla ^{2}c,
\label{equ:PRR-convdiff}
\end{equation}
in a passive scalar in the fluid with dimensionless concentration
\textit{c} having a diffusion coefficient $\Gamma $\textit{.} At
moment $t=0$, $c=0$ everywhere in the fluid while $c=1$ is maintained
at the fluid-solid interface. This way, scalar is transferred to the
fluid: for each particle \textit{i} at a rate $\dot{m}_{i}=-\Gamma
\int _{A_{i}}\partial c/\partial ndA$ with $A_{i}$ the surface of the
particle and \textit{n} the particle-outward normal. The transfer
process is parameterized through a transfer coefficient $k_{i}$
according to $\dot{m}_{i}=k_{i}S\Delta c$ with $S=\pi d_{p}^{2}$ and
$\Delta c$ a ``characteristic'' concentration difference (specified
below). The Sherwood number is defined as
${Sh}_{i}=k_{i}d_{p}/\Gamma $ and (finally) the average Sherwood
number is the average over all \textit{N} particles in the domain
${Sh}=\frac{1}{N}\sum _{i=1}^{N}{Sh}_{i}$.  

\begin{figure}
  \centering
  \includegraphics[width=.4\linewidth]
  {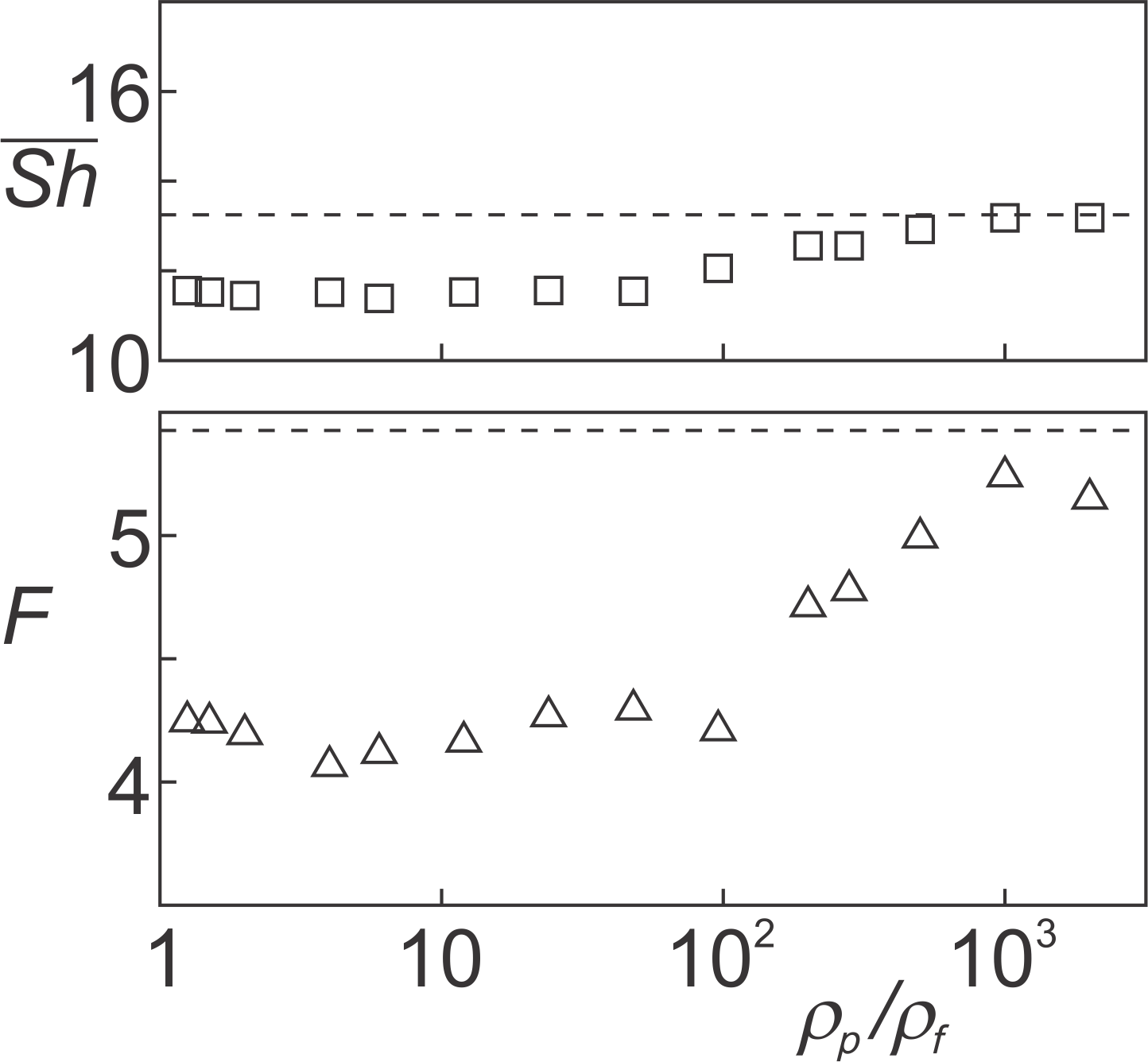}
  \caption{
    Average Sherwood number $\overline{{Sh}}$ (top) and drag
    \textit{F} (bottom) as a function of density ratio $\rho _{p}/\rho_{f}
    $ in a fully periodic domain with freely moving particles with
    overall \revision{} {particle} volume fraction $\left\langle \phi \right\rangle
    $=0.2 and Schmidt number ${Sc}=\nu /\Gamma =300$. Dashed
    lines indicate results of a fixed-particle simulation.} 
  \label{fig:PRR-shwd}
\end{figure}

This average Sherwood number depends on time with
${Sh}\rightarrow \infty $ at $t=0$ when scalar penetrates
rapidly into fresh fluid. If, however, we define $\Delta
c=1-\left\langle c\right\rangle $ with $\left\langle c\right\rangle $
the (time dependent) fluid volume-average concentration, ${Sh}$ is
approximately constant over a wide time window \cite{derksen2014};
this constant value we denote as $\overline{{Sh}}$. It now is
interesting to note that $\overline{{Sh}}$ depends on the
solids over liquid density ratio $\rho _{p}/\rho_{f} $, see
Figure~\ref{fig:PRR-shwd} with $\overline{{Sh}}$ increasing with
increasing $\rho _{P}/\rho_{f} $ and reaching a plateau if $\rho _{P}/\rho_{f}
\approx 10^{3}$. The plateau value is approximately equal to
$\overline{{Sh}}$ obtained from a fixed-particle simulation. 

Figure~\ref{fig:PRR-shwd}  also shows drag force data as a function of
$\rho _{P}/\rho_{f} $ and we see that the transition of \textit{F} occurs
in the same density ratio range as $\overline{{Sh}}$ lending
credit to the hypothesis that clustering in moderate Stokes number
suspensions leads to a reduction in mass transfer. 

\begin{figure}
  \centering
  \includegraphics[width=.6\linewidth]
  {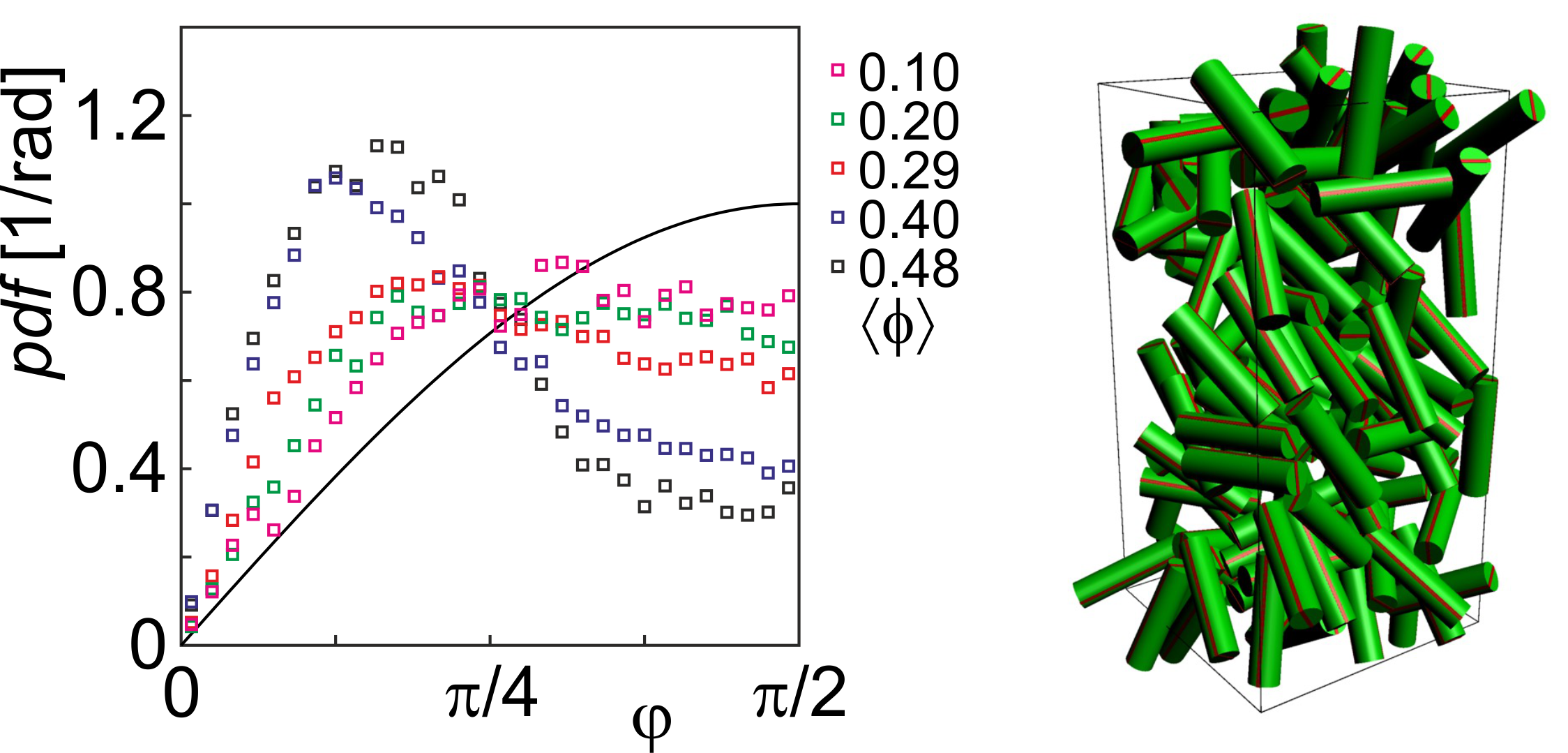}
  \caption{
    Left: orientation distribution of fluidized cylinders with aspect
    ratio $\mathrm{\ell }/d$=4 in terms of their angle $\varphi $ with
    the vertical for a range of overall \revision{}{particle} volume fractions
    $\left\langle \phi \right\rangle $ as indicated. The drawn $\sin
    \varphi $ curve is an isotropic orientation distribution. Right:
    an impression for $\left\langle \phi \right\rangle $=0.29. Density
    ratio $\rho _{p}/\rho_{f} $=2 and ${Ga}$=29 
    \cite{derksen2019liquid}.}
  \label{fig:PRR-cylinders}
\end{figure}

We now briefly \revision{}{turn} to the topic of non-spherical particles. To the
best of the authors’ knowledge, the problem of fluidization of many
non-spherical particles computed with PR-DNS has not been considered in
the literature with any other shape than finite-size \revision{}{rods} in
\cite{derksen2019liquid}.
Among other things, 
\cite{derksen2019liquid} shows -- next to
preferential concentration -- also preferential orientation and the way
it depends on the \revision{}{particle} volume fraction $\phi $
{and the length over diameter ratio} $\mathrm{\ell
}/d$ of the cylinders. Figure~\ref{fig:PRR-cylinders} shows a sample
result for $\mathrm{\ell }/d$=4 in terms of the distribution of the
angles $\varphi $ of the fluidized cylinders with the vertical. A
fully isotropic orientation distribution implies a distribution of
$\varphi $ according to $\sin \varphi $. For $\phi \leq 0.29$ the
orientation distribution is approximately isotropic. For $\phi \geq
0.40$ the cylinders have a clear preference for small angles $\varphi
$.  

The above results were all obtained through lattice-Boltzmann
simulations (see Chapter~5) with an immersed boundary method for
achieving no-slip at the solid-liquid interfaces.

Contributions of Euler-Lagrange simulations of fluidization with
non-spherical particles  are more numerous
\cite{vollmari2016experimental,gao2021development} as compared to
particle-resolved simulations. More generally, the complementary
problem to fluidization, the free settling of non-spherical particles,
has not been examined much with PR-DNS either,
\cite{seyed:21} %
with cubes and
\cite{fornari:18,frohlich2021particle}
with spheroids being
a few exceptions. Instead, researchers have addressed other,
presumably simpler, problems as preliminary steps to pave the way to
the full PR-DNS of freely settling or fluidized non-spherical
particles. These other problems include (i) the flow past a
non-spherical fixed particle to gain insight into the shape and
angular position dependent drag, lift and torque coefficients
(\cite{Holzer2009,Zastawny2012,richter2012drag,richter2013new,sanjeevi2017orientational,frohlich2020correlations}
with a spheroid,
\cite{Holzer2009,vakil2009drag,pierson2019inertial,kharrouba2021flow}
with a finite-size cylinder and
\cite{saha2004three,Holzer2009,meng2021wake} with a polyhedron), (ii)
the flow past a random array of non-spherical fixed particles to
understand drag, lift and torque coefficient modulations through
neighboring particle flow disturbances (\cite{li2021effect} with
spheroids, \cite{tavassoli2015direct,sanjeevi2020hydrodynamic} with
spherocylinders and \cite{chen2018development} with polyhedrons), and
(iii) the free settling of a single non-spherical particle to
investigate path trajectories and wake instabilities
(\cite{moriche2021single} with a spheroid, \cite{Chrust2013} with a
finite-size cylinder and \cite{rahmani2014,seyed2019dynamics} with a
polyhedron). The PR-DNS of flows laden with non-spherical particles
require more spatial resolution than the counterpart with spheres, and
therefore larger computing resources, while the problem of collisions
of non-spherical particles is addressed at length in
Chap~7. 
The analysis of the flow might also be \revision{}{slightly} more challenging with
non-spherical particles than with spheres, but we believe that the PR-DNS
of flows laden with non-spherical particles is primarily a high
performance computing problem. With the increasing power offered by
supercomputers, we expect to see more contributions to the literature
in the near future on this topic. 

\section{PR-DNS of unbounded \revision{flow with suspended particles}{flows in the dilute regime}}
\label{sec:PRR-unbounded}
In the present section we consider the collective dynamics of particles suspended in a body of fluid far from any solid boundaries, \revision{}{focusing on the dilute regime}. In a first part we will focus on systems where the fluid is initially at rest\revision{}{,} and in a second part on configurations where the background flow corresponds to the classical homogeneous isotropic turbulence.
\subsection{Settling in initially ambient fluid}
\label{sec:PRR-unbounded-ambient}
The behavior of a single particle settling in ambient fluid is a surprisingly rich topic, as a variety of path regimes arise in different parts of the parameter space (density ratio, Galileo number, particle shape), with the single particle settling e.g.\ straight, vertically, on a straight oblique path, with time-periodic oscillations, in zig-zag or chaotically.
This topic has been exhaustively studied in the past, either experimentally or numerically, in the latter case typically with the aid of interface-conforming methods (cf.\ the review by \citep{ern:12}).

When a collective of particles is considered, the nature of the interaction becomes even more complex and is \revision{nowadays}{} still an object of active research.
The most striking features concern the formation of clusters due to
particle interaction and the possible increase or decrease of the
settling velocity, which requires to be properly defined.
In this case, the settling velocity $\revision{W}{W_{rel}}^{(i)}$
of the $i^{th}$ particle corresponds to the relative velocity
between a particle and the fluid velocity averaged in the entire
domain, and reads (supposing that gravity acts in the negative
\revision{z}{\textit{z}}-direction): 
\begin{eqnarray}
  W\revision{}{_{rel}}^{(i)}(t)
  &=&
      V_{p,z}^{(i)}(t)-\lab u_{f,z}
      \rab_{{\cal V}_f}(t)\,
      \label{eq:PRR-global_W}
\end{eqnarray}
where 
\revision{}{$\lab \textbf{u}_{f}
  \rab_{{\cal V}_f}$ is the fluid velocity averaged over the
  volume occupied by the fluid}.
$\revision{W}{W_{rel}}$ is then the value obtained once ensemble
averaging (\ref{eq:PRR-global_W}) over all particles, 
and $W_s$ henceforth denotes the terminal velocity of an isolated particle.

\begin{figure}
  {\small
    \begin{tabular}{cl}
      \ACcircfull&$\phi=0.01$, $\revision{Ga}{\mathrm{Ga}}=145$
                   \cite{fornari:18a}\\
      \ACdiamond& $\phi=0.05$, $19 \leq \revision{Ga}{\mathrm{Ga}}
                  \leq 200$ \cite{fornari:16d}\\
      \ACtrileft& oblate spheroids (aspect ratio 1/3),
                  $\revision{Ga}{\mathrm{Ga}}=60$ \cite{fornari:18b}\\
      \textcolor{ACblue}{\ACcirc}& spheres,
                          \textcolor{ACblue}{\ACsquare} cubes,
                           $\revision{Ga}{\mathrm{Ga}}=160$ \cite{seyed:21}\\
      \textcolor{ACdarkblue}{\ACcirc}& spheres,
                          \textcolor{ACdarkblue}{\ACsquare} cubes,
                           $\revision{Ga}{\mathrm{Ga}}=70$
                                       \cite{seyed:21}\\
    \end{tabular}
    \begin{tabular}{cl}
      \textcolor{ACred}{\ACtriup}&
                                   $0.05 \leq \phi \leq0.4$, 
                                   $\revision{Re_{ps}}{\mathrm{Re_{ps}}}
                                   < 10^{-3}$ \cite{nicolai:95}\\
      \textcolor{blue}{\ACtridown}& $2\times 10^{-4} \leq \phi \leq 0.001$,
                                    $110 \leq \revision{Ga}{\mathrm{Ga}} \leq 310$ 
                                    \cite{huisman:16}\\
      \textcolor{ACgreen}{\ACdiamond}& $\phi=0.005$, $\revision{Ga}{\mathrm{Ga}}=121$,
                          \textcolor{ACgreen}{\ACcirc} $\revision{Ga}{\mathrm{Ga}}=178$,
                                       \cite{uhlmann:14a}\\
      \textcolor{ACgreen}{\ACcircfull}& $\phi=0.0005$, $\revision{Ga}{\mathrm{Ga}}=178$,
                                        \cite{todor_phd}\\
      \textcolor{ACyellow}{\ACcross}& $\phi=0.003$, 
                     \textcolor{ACyellow}{\ACcirc}\, $\phi=0.01$, $0.1
                                      \leq
                                      \revision{Re_{ps}}{\mathrm{Re_{ps}}}
                                      \leq 10$ \cite{climent:03}\\
      \textcolor{ACpurple}{\ACcross}& $\phi=0.003$, 
                     \textcolor{ACpurple}{\ACcirc}\, $\phi=0.01$, $1
                                      \leq
                                      \revision{Re_{ps}}{\mathrm{Re_{ps}}}
                                      \leq 300$ \cite{zaidi:14}\\
    \end{tabular}
  }
  \\[0ex]
  \begin{minipage}{2ex}
    \rotatebox{90}{$\proofedit{w_{rel}}{W_{rel}}/W_{s}$}
  \end{minipage}
  \begin{minipage}{0.5\linewidth}
    \centerline{$(a)$}
    \includegraphics[width=0.9\linewidth]
    {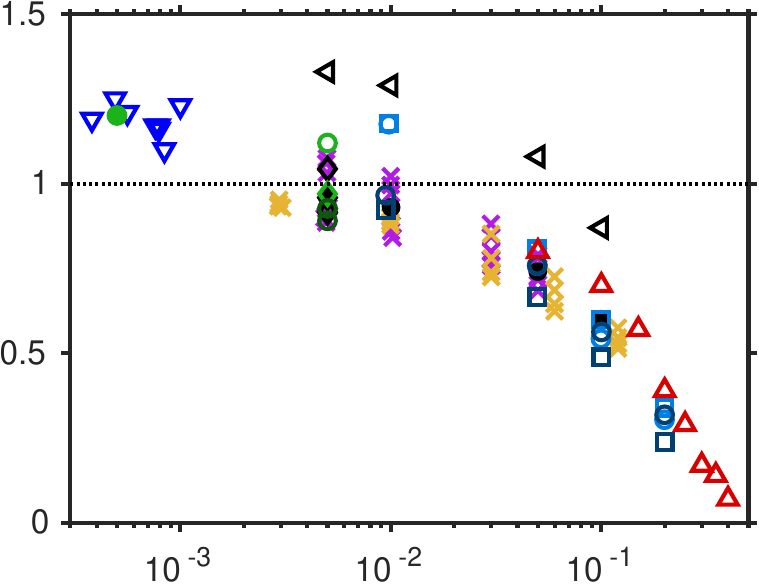}
    \centerline{$\phi$}
  \end{minipage}
  \begin{minipage}{0.5\linewidth}
    \centerline{$(b)$}
    \includegraphics[width=0.9\linewidth]
    {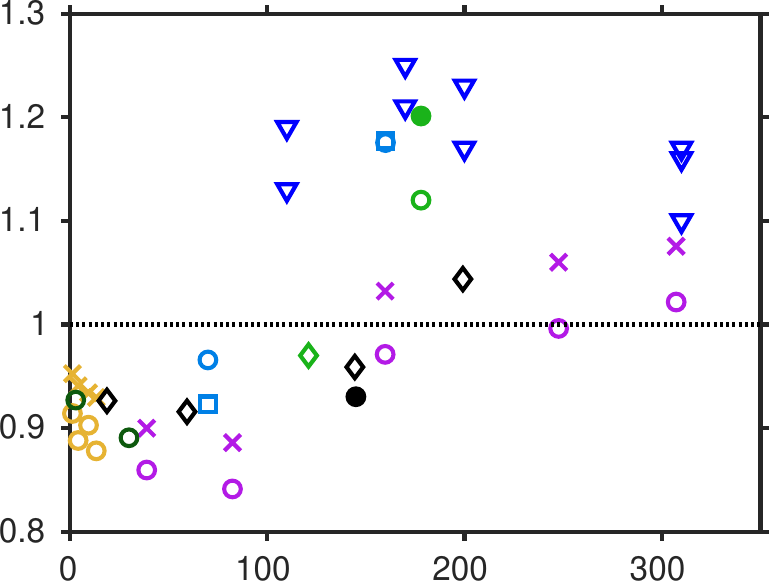}
    \centerline{$\revision{Ga}{\mathrm{Ga}}$}
  \end{minipage}
  \caption{Evolution of the mean particle settling velocity
     $W\revision{}{_{rel}}$ scaled by the corresponding single-sphere
     value $W_s$ for different simulations and experiments in ambient
     fluid represented as a function of the solid volume fraction (a)
     and Galileo number (b). When not mentioned in the legend, the
     data refers to spherical particles. \revision{}{Colorstyle is
       common for (a) and (b), data from references \cite{climent:03}
       and \cite{zaidi:14} are represented exclusively with crosses on
       (a).}} 
   \label{fig:PRS-settling_vel_ambient}
\end{figure}

Figure \ref{fig:PRS-settling_vel_ambient} gathers results from the literature on the evolution of the settling velocity at different \revision{solid}{particle} volume fraction and Galileo numbers.
The settling of many spheres in an initially ambient fluid has been first numerically investigated with DNS by \citet{kajishima:02} for single sphere particle Reynolds number $\revision{Re_{ps}}{\mathrm{Re_{ps}}}$ ranging from 50 to 400 \revision{}{and density ratio $\rho_p/\rho_f=8.8$}. They used periodic boundary conditions in the three directions and neglected particle rotation. They observed the formation of columnar clusters for $\revision{Re_{ps}}{\mathrm{Re_{ps}}}$ larger than 100 and a global decrease of the drag coefficient, attributed to wake interactions, with the effect to globally increase the settling velocity.
Later \citet{kajishima:04b} explored similar configurations while accounting for particle rotation and considering larger computational setups at $\revision{Re_{ps}}{\mathrm{Re_{ps}}} \approx 300$. This work confirmed the formation of columnar clusters with rotational particles with a slightly smaller increase in settling velocity.
\citet{uhlmann:14a} later focused on the influence of the settling regime. They considered two configurations featuring the same solid volume fraction ($\phi = 5 \,\times\,10^{-3}$) but two different single-sphere settling regimes: a steady axi-symmetric regime with $\revision{Ga}{\mathrm{Ga}}=121$ and a steady oblique regime with $\revision{Ga}{\mathrm{Ga}}=178$. They showed that particles tend to cluster into columns only for the the steady oblique regime, leading to a global increase of the settling velocity.
The experimental work of \citet{huisman:16} has confirmed that the cluster formation into columns does not constitute a numerical artifact. They investigated Galileo numbers ranging from $110$ to $310$ and showed that for $\revision{Ga}{\mathrm{Ga}} \geq 170$ particles tend to align vertically, while the cluster formation stays relatively weak at $\revision{Ga}{\mathrm{Ga}}=110$.
The level of clustering was quantified with the aid of Vorono\"i
tessellation analysis, as introduced by \citet{monchaux:10}, and
described in
chapter~2. 
Such analysis has been previously used in
\revision{\cite{uhlmann:14a}}{the work of \citet{uhlmann:14a}} for
finite-size particles. Note that it was shown
in \citep{uhlmann:20}
that great care has to be taken in the analysis to account for finite size effects, since the constraint of non-overlapping particles and the number of particles both affect the reference of randomly positioned particles.
\revision{}{It is also necessary to consider large computational
  domains as it might affect the formation of clusters.
} 

The picture is different when the \revision{solid}{particle} volume fraction increases or when the Galileo number decreases as represented on figure 
\ref{fig:PRS-settling_vel_ambient}.
\citet{climent:03} and \citet{zaidi:14} numerically observed a decrease of the settling velocity for these cases, with a similar trend as experimentally shown by \citet{nicolai:95} at very small settling Reynolds numbers ($\revision{Re_{ps}}{\mathrm{Re_{ps}}} \leq 10^{-3}$).
This velocity decrease is usually attributed to the fact that settling generates an upward flow to compensate for the particle downward motion and ensure zero mixture velocity. This phenomenon is known as the hindrance effect for which \citet{richardson:97} proposed an approximation  by a power law of type
\begin{eqnarray}
  {\revision{W}{W_{rel}}}/{W_s} &=& (1-\phi)^{n},
   \label{eq:PRR-RichardsonZaki_correlation}
\end{eqnarray}
based on experimental data.
In \cite{fornari:16d} a set of Galileo numbers ranging from $19$ to $200$ with density ratios close to unity (ranging from $1.0035$ to $1.038$) was investigated and a global decrease of $\revision{W}{W_{rel}}$ compared to $W_s$ was observed for particles settling in the steady axi-symmetric regime, while particles at $\revision{Ga}{\mathrm{Ga}}=200$ settle with a higher velocity. 
Later \citet{fornari:18a} focused on the case $\revision{Ga}{\mathrm{Ga}}=145$ with different \revision{solid}{particle} volume fractions and observed a decrease of $\revision{W}{W_{rel}}$ with $\phi$ which can reach $60\%$ of its single-particle value at $\phi=1\%$. 

The recent numerical work of \citet{seyed:21} have shown that this transition from an increase of the settling velocity at very low \revision{solid}{particle} volume fraction to hindrance effect at larger $\phi$ seems to take place for spheres and cubes at $\revision{Ga}{\mathrm{Ga}}=160$, while spheres and cubes falling at $\revision{Ga}{\mathrm{Ga}}=70$ only show a decrease of $\revision{W}{W_{rel}}$.
Here again the settling regime of a single particle seems to play a significant role since sphere and cube respectively fall in the steady oblique and helical regime at $\revision{Ga}{\mathrm{Ga}}=160$ while both sphere and cube follow a vertical path at $\revision{Ga}{\mathrm{Ga}}=70$.

One of the major advantages of numerical simulation is that it gives us access to the velocity field in the vicinity of each particle, opening with it the discussion on the flow conditions actually experienced by each particle. This leads to a distinction between two definitions of the settling velocity. As mentioned above, the first definition $\revision{W}{W_{rel}}$ considers the volume averaged velocity $\lab \bu_f(t) \rab_{{\cal V}_f}$ to be relevant for the fluid phase. But here the flow is not uniform, and the formation of clusters is expected to create large zones of downward fluid motion in clusters and upward motion in low-concentration regions. 
\citet{Kidanemariam2013} %
introduced another definition in which the velocity relevant for the fluid phase corresponds to the fluid velocity $\bu_f^{{\cal S}_i}(t)$ averaged on a spherical shell centered at the particle position. 
Tests on the influence of the shell diameter $d_p^{{\cal S}}$ \revision{}{performed for a uniform unbounded flow past a sphere }indicated that a radius of $1.5d_p$ gives the best compromise \revision{}{\cite{Kidanemariam2013}}.
Indeed, if $d_p^{{\cal S}}$ is too small then the mean velocity is dominated by the wake of the corresponding particle while the velocity computed with a large $d_p^{{\cal S}}$ would not be representative of the flow actually seen by the particle.
Due to finite-size effects the system is strongly two-way coupled and
inhomogeneous at the particle scale so that this shell-averaged
velocity is different from the "undisturbed" and the "free stream"
velocities as discussed in chapter~7. 
This velocity is also conceptually different from the
filtered velocity used in methods relying on volume-filtered field
(see for instance chapter~11) 
as the average is only computed on a shell to minimize self-disturbance contributions.
The corresponding settling velocity is defined as follows \revision{}{for a given particle}
\begin{eqnarray}
   W\revision{}{_{rel}}^{{\cal S}_{(i)}}(t)
      &=&
      V_{p,z}^{(i)}(t)-u_{f,z}^{{\cal S}_i}(t)\,\revision{.}{,}
   \label{eq:PRR-local_Rep}
\end{eqnarray}
\revision{}{and $W_{rel}^{{\cal S}}(t)$ refers to the velocity
  obtained once ensemble averaging (\ref{eq:PRR-local_Rep}) over the
  particles.} 
\citet{uhlmann:14a} observed that the particle-averaged settling velocity $\revision{W}{W_{rel}}^{{\cal S}}(t)$ is comparable to the corresponding single particle value $W_s$ with and without cluster formation.
As the fluid velocity within clusters is mostly directed downwards, the enhancement of $\revision{W}{W_{rel}}$ could simply be the effect of preferential sampling of the flow.
The experimental work of \citet{huisman:16} confirms this trend since particles with small Vorono\"i volumes are shown to fall with an increased Reynolds number while particles with \revision{small}{large} volumes tend to settle with a decreased velocity. This decrease of velocity in low density regions can be attributed to the formation of upward flow regions outside of clusters.

Less is known on the properties of the pseudo-turbulence generated by
the particles while they settle, and the detailed characteristics of
these fluctuations are nowadays still the object of
speculations. Numerical work of \citet{kajishima:02} and
\citet{todor_phd} indicates that this turbulence displays similar
properties as turbulence generated by the rise of bubbles in a
quiescent flow, which has been described by
\citet{risso:11}. \citet{riboux:09} have already shown experimentally
that a random set of fixed spheres generates turbulence featuring the
same structure as a rising bubble cloud. The turbulence as described
by \revision{\cite{risso:11}}{\citet{risso:11}} results in the superposition of Gaussian wake
contributions, leading to a turbulence energy spectrum featuring a
$-3$ power law. This type of energy spectra has been numerically
observed \cite{todor_phd}, which tends to indicate that similar wake
influence is at play. 
\begin{figure}
\begin{minipage}{0.325\linewidth}
	\begin{overpic}[width=1.\linewidth]
		{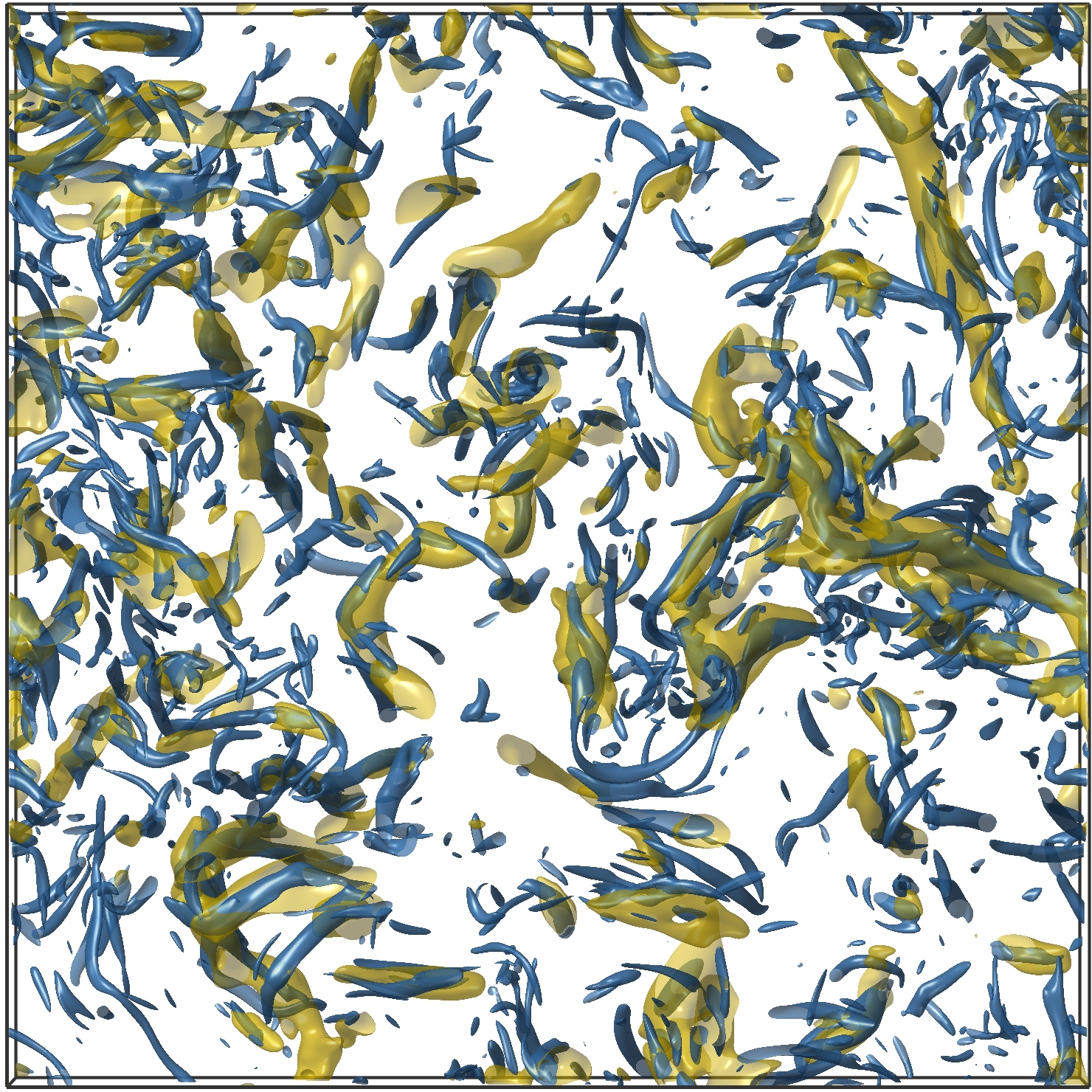}
	\end{overpic}
	\begin{overpic}[width=1.\linewidth]
		{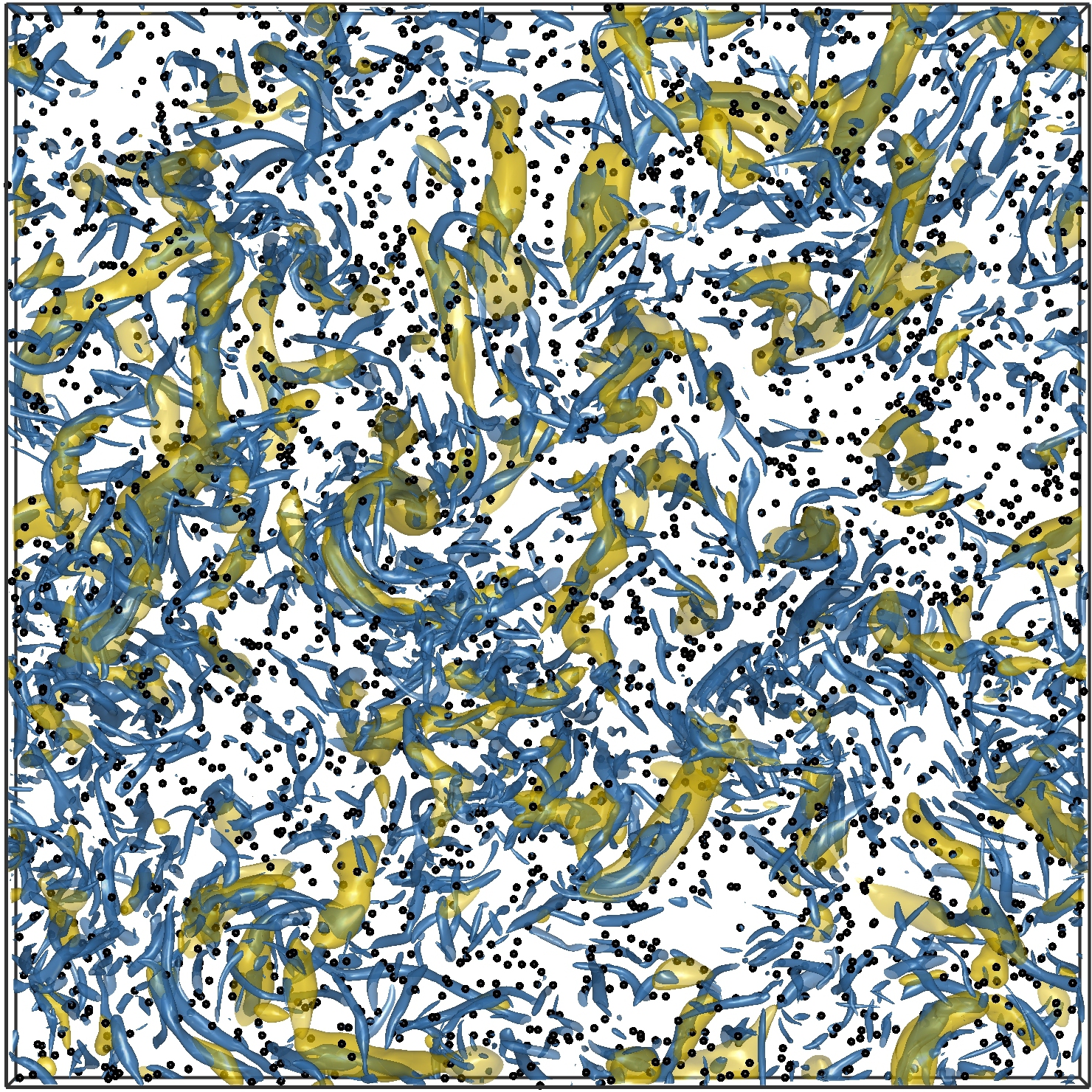}
	\end{overpic}
\end{minipage}
\begin{minipage}{0.325\linewidth}
	\begin{overpic}[width=1.005\linewidth]
		{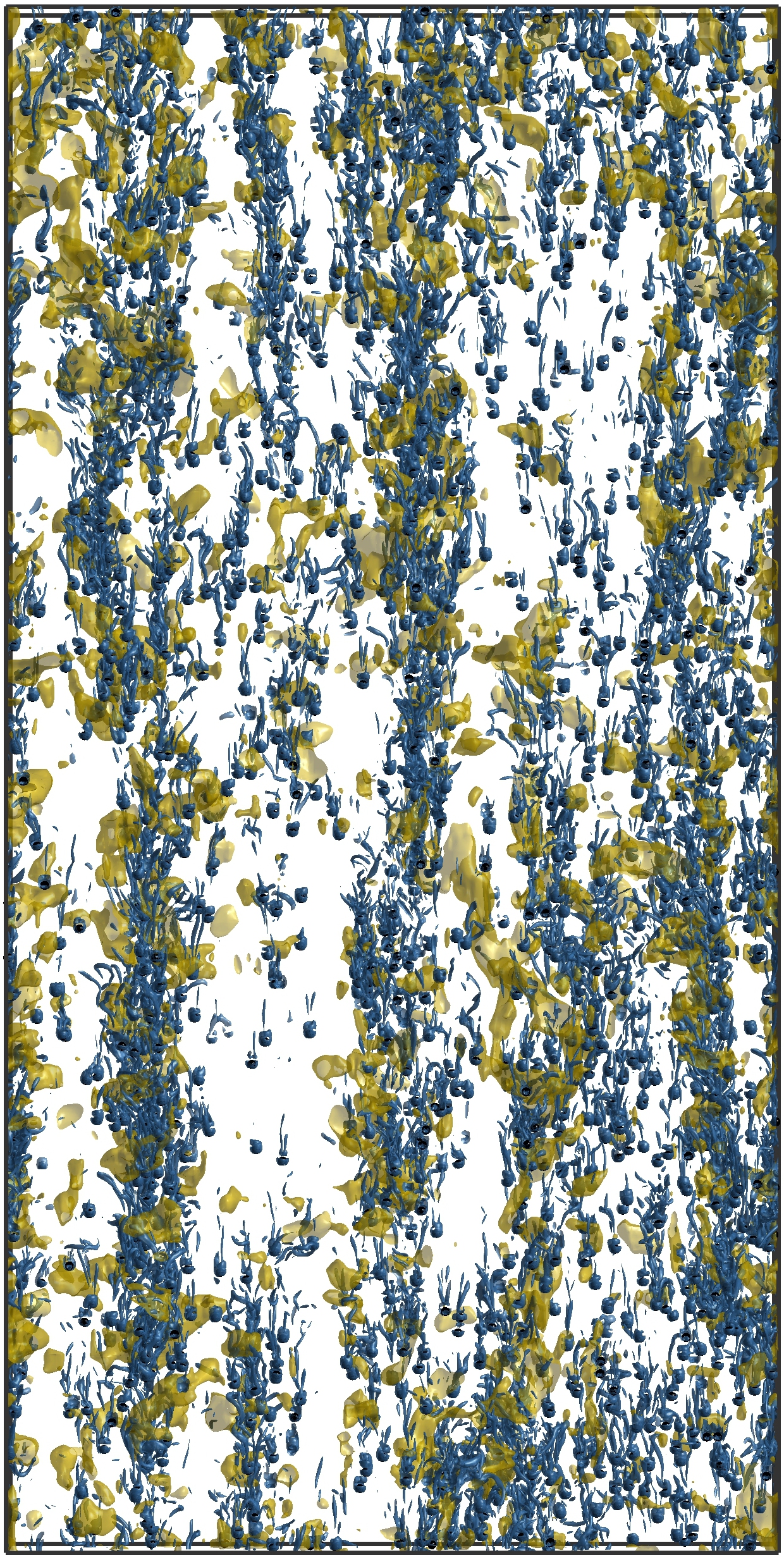}
	\end{overpic}
\end{minipage}
\begin{minipage}{0.325\linewidth}
	\begin{overpic}[width=1.005\linewidth]
		{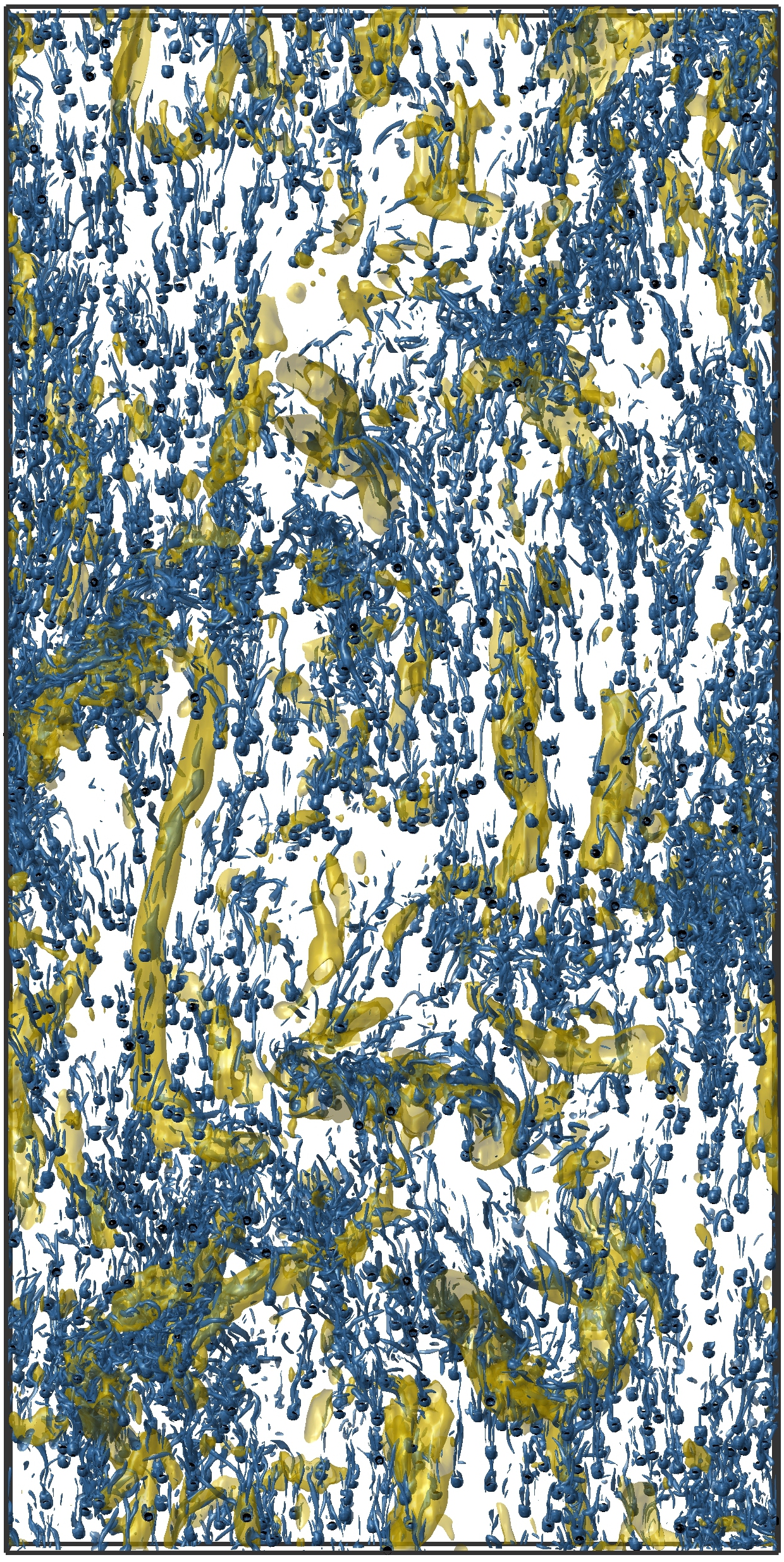}
	\end{overpic}
\end{minipage}
\caption{Visualization of the flow structure taken from
  \citet{chouippe:18}: Isocontour of the $Q$ criterion for the
  unfiltered field (blue) and the field box-filtered at a scale
  $\Delta_{filt}$ (yellow) for a single-phase case at $Re_\lambda=95$
  (top left), for $Ga=0$ and $Re_\lambda=120$ (bottom left,
  $\Delta_{filt}=3.7d_p$), $Ga=178$ and ambient flow (center,
  $\Delta_{filt}=5.6d_p$), $Ga=180$ and $Re_\lambda=142$ (right,
  $\Delta_{filt}=5.6d_p$). The visualizations represent the total
  domain in the vertical plane and one eighth of the domain in the
  horizontal plane (into the page).} 
\label{fig:PRS-3D_visus}
\end{figure}
\subsection{Finite-size particles in turbulent background
  flow} 
\label{sec:PRR-unbounded-turb}
We consider now configurations where the carrying flow is still unbounded but turbulent, focusing on the idealized configuration of homogeneous-isotropic-turbulence as represented on figure \ref{fig:PRS-3D_visus}, where the typical worm-like turbulent structures are made visible by representing iso-surfaces of the Q-criterion. 
Such flows are usually generated in simulations by the use of a forcing term in the momentum equation of the Navier-Stokes equations, and mainly two types of schemes have been used in the literature: it either depends on flow characteristics such as velocity field or dissipation rate, or it can be independent of the flow field. The second category is usually preferable in the presence of finite-size particles, as the system might become unstable due to two-way coupling effects.
Most of these external forcings are formulated in spectral space and consist in randomly injecting energy in the largest eddies of the system \cite{eswaran:88}.
In the absence of gravity ($\revision{Ga}{\mathrm{Ga}}=0$), the main questions addressed in the literature mostly concern the possible formation of particle clusters, and scaling of particle acceleration and velocity, focusing on the deviation from the point-particle limit.
\citet{uhlmann:16a} have numerically investigated configurations featuring two different Reynolds numbers ($\revision{Re_\lambda}{\mathrm{Re_\lambda}}= 115$ and $140$) with particles larger than the Kolmogorov lengthscale ($d_p/\eta=5,\,11$) and slightly denser than the fluid ($\rho_p/\rho_f=1.5$), in the dilute regime ($\phi=0.005$).
The corresponding Stokes numbers are larger than unity if based on the Kolmogorov time-scales ($\revision{St_\eta}{\mathrm{St_\eta}} = 2.5,\,10.7$) and smaller than unity if based on the large-eddy turnover time ($\revision{St_T}{\mathrm{St_T}}=0.06,\,0.29$).
With the aid of Vorono\"i tessellation analysis a weak formation of clusters was shown \cite{uhlmann:16a} and the authors analyzed particle position with respect to the position of 'sticky-points' \cite{goto:08}.
Those points verify $\mathbf{e}_1\cdot\mathbf{a}_f=0$ and $\lambda_1 >
0$, where $\mathbf{a}_f$ denotes the fluid acceleration, $\lambda_1$
is the largest eigenvalue and $\mathbf{e}_1$ the corresponding
eigenvector of the symmetric part of the acceleration gradient tensor
(cf.\ also chapter~2). 
The analysis shows a small but statistically significant increase of the probability of finding `sticky-points' in the particles' vicinity, implying that a clustering mechanism similar to the sweep-stick mechanism \cite{goto:08} might be at play.
\revision{This observation is also consistent with the generalized
  sweep-stick mechanism \cite{oka:21}, which shows that inertial
  point-particles can also form clusters under the influence of
  turbulence eddies larger than the Kolmogorov length-scale, meaning
  that the Stokes number $\revision{St}{\mathrm{St}}_\eta$ is not
  necessarily the best indicator to predict preferential
  accumulation.}{%
  Note that the authors in \cite{uhlmann:16a} have not found any
  significant direct evidence of the centrifugal mechanism
  \citep{maxey:87} at work. 
}

Turning now to the evolution of particle velocity and acceleration, the shapes of their p.d.f. seem to be relatively well established in the literature but not their variance.
The normalized p.d.f. of single velocity components appears to be Gaussian as the carrying fluid 
\citep{uhlmann:16a, homann:10, yeo:10}, and the p.d.f. of acceleration has been shown to follow a universal shape related to a lognormal distribution of the magnitude of the particle acceleration \citep{mordant:04,qureshi:07}.
The relative difference between the particle and fluid velocity variance has been first shown to follow a $d_p^{2/3}$ power law \cite{homann:10} for $d_p \gtrsim 5\eta$, which has been later also observed by \citet{uhlmann:16a} for $d_p \gtrsim 3\eta$.
Different trends have been observed for the acceleration variance: experimental work in Von K\'arm\'an flow \citep{voth:02,brown:09,volk:11} or in a wind tunnel \citep{qureshi:07} exhibited a $(d_p/\eta)^{-2/3}$ or $(d_p/\eta)^{-0.81}$ power-law at large flow Reynolds number, while DNS performed at lower $\revision{Re_\lambda}{\mathrm{Re_\lambda}}$ exhibits an evolution proportional to  $(d_p/\eta)^{-4/3}$ \citep{homann:10,cisse:15,uhlmann:16a}. 
A $(d_p/\eta)^{-2/3}$ power law can be attributed to the fact that particle acceleration is mostly affected by flow scales of the order of $d_p$, with a correction accounting for intermittency leading to the $(d_p/\eta)^{-0.81}$ power-law, while an evolution in $(d_p/\eta)^{-4/3}$ seems to be the effect of the largest scale that predominantly affect particle motion through sweeping mechanisms.

In the absence of gravity and in the dilute regime, particles have been shown to have little influence on the flow statistics \citep{uhlmann:16a} so that the flow can globally be considered as nearly one-way coupled even for particles larger than $\eta$.
But this is not the case anymore in the presence of gravity, and an important parameter in that case is the relative turbulence intensity $\revision{I}{\mathrm{I}}=u_{rms}/W_s$, namely the ratio between the characteristic velocity of the background forced turbulence and the settling velocity of an isolated particle in quiescent flow.
This can be explained by considering the global energy balance: for
this we introduce the instantaneous kinetic energy
$E_k(t)=\mathbf{u}\cdot\mathbf{u}/2$ (where $\mathbf{u}$ refers to the
composite velocity field defined throughout the joint volume occupied
by either the fluid or the particles) and the time rate of 
change of its volume average \revision{}{\citep{chouippe:15a}:}
\begin{eqnarray}\label{eq:PRR-kinetic-energy-balance}
   \frac{\mathrm{d}\left< E_k\right>_{{\cal V}_\infty}}{\mathrm{d}t}
   &=&
      -\mathlarger{\varepsilon}_{{\cal V}_\infty} + \psi^{(turb)} + \psi^{(p)}\,
\end{eqnarray}
where $\mathlarger{\varepsilon}_{{\cal V}_\infty}$ refers to the
dissipation rate averaged over the entire volume of the system, and
$\psi^{(turb)}$ and $\psi^{(p)}$ to the work done by the turbulence
forcing used to generate the turbulent background flow and the
fluid-particle coupling, respectively.
\revision{}{%
  Note that various authors have used alternative forms of the kinetic
  energy budget in the context of PR-DNS
  \citep[e.g.][]{mehrabadi:16,schneiders:17}.}
\revision{In \cite{chouippe:18} the settling of particles at
  $\revision{Ga}{\mathrm{Ga}} \approx 180$ in two configurations
  featuring $d_p/\eta = 6.8 - 8.5$,
  $\revision{Re_\lambda}{\mathrm{Re_\lambda}}=95 - 142.2$, and
  turbulence intensities $I=0.14 - 0.22$ has been
  investigated.}{\citet{chouippe:18} have investigated the settling of
  particles at $\revision{Ga}{\mathrm{Ga}} \approx 180$ in two
  configurations featuring $d_p/\eta = 6.8 - 8.5$,
  $\revision{Re_\lambda}{\mathrm{Re_\lambda}}=95 - 142.2$, and
  turbulence intensities $\mathrm{I}=0.14 - 0.22$.} They observed that
the work done by the turbulence forcing is barely affected by the
presence of the particles and that $\psi^{(p)}$ is dominated by a term
$\psi_{(pot)}^{(p)}$ which
\revision{}{arises from the slip velocity
  between particles and the fluid and}
is defined as
\begin{eqnarray}
   \frac{\psi_{(pot)}^{(p)}}{\mathlarger{\varepsilon}}
      &=&
      \phi \revision{Ga}{\mathrm{Ga}}^3 \left(\frac{d_p}{\eta} \right)^{-4} 
      \frac{\revision{V_{p,z}}{\lab V_{p,z}^{(i)} \rab}-u_{f,{\cal V}_\infty}}{u_g}\, ,
      \label{eq:PRR-psi_pot_details}
\end{eqnarray}
meaning that the dissipation tends to increase under the release of the particles' potential energy while they settle.
A first rough estimation of this increase can be obtained based on equation (\ref{eq:PRR-psi_pot_details}) under the approximation $\revision{V_{p,z}}{\lab V_{p,z}^{(i)}\rab}-u_{f,{\cal V}_\infty} \approx W_s$. It emphasizes the importance of the particle size, \revision{solid}{particle} volume fraction and Galileo number on the modification of the background flow.
Such an approximation would then predict that at a given Galileo number the two-phase dissipation would linearly increase with $\phi$, which is in accordance with the numerical observation of \citet{fornari:18a} at $\revision{Ga}{\mathrm{Ga}}=145$, $d_p/\eta=12$ and $\revision{Re_\lambda}{\mathrm{Re_\lambda}}=90$. \revision{}{This importance of the contribution associated to slip velocity is also qualitatively in accordance with pseudo-turbulent configurations \cite{mehrabadi:15}}.
\revision{}{The influence of particle clusters on the background turbulence is difficult to isolate as the level of clustering appears to be influenced by the turbulence intensity.}

Concerning now the influence of turbulence on particle settling and potential clustering it seems that homogeneous-isotropic turbulence at modest Reynolds numbers acting on particles larger than the Kolmogorov scale globally tends to decrease the level of clustering and the settling velocity compared to equivalent configuration with initially ambient flows \citep{chouippe:18, fornari:16a, fornari:16d, fornari:18a}. \citet{chouippe:18} have shown, based on Vorono\"i tessellation analysis, that forcing turbulence rapidly disturbs the wake-induced columns without completely destroying them, but it does not display a monotonic trend with turbulence intensity. The relative size of the particles with respect to the scales of the forced turbulence seems to play an important role, but the mechanisms at play are still the object of an ongoing discussion, and not all parameters have been systematically swept.

\begin{figure}
  \centering
  \begin{minipage}{2.5ex}
    \rotatebox{90}{%
      $\proofedit{w_{rel}}{W_{rel}}/W_s$ 
    }
  \end{minipage}
  \begin{minipage}{.45\linewidth}
    \includegraphics[width=\linewidth]
    {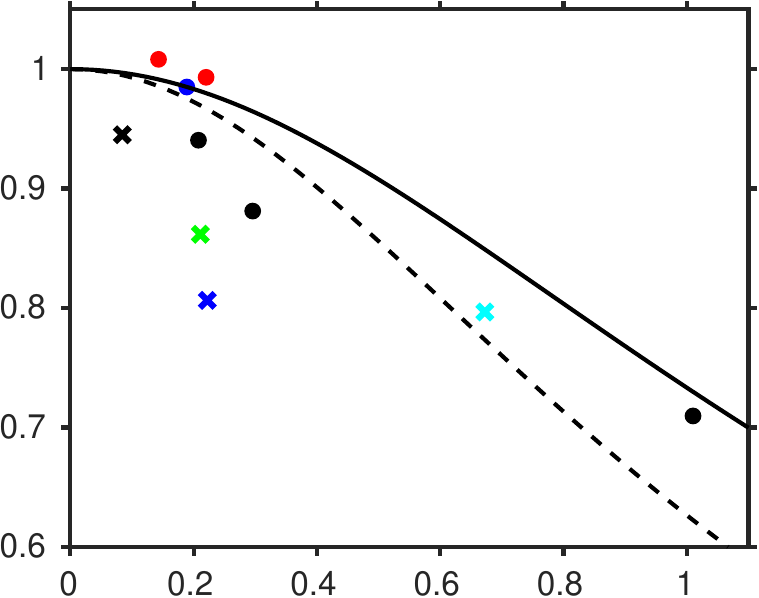}
    \\
    \centerline{$\revision{I}{\mathrm{I}}$, $\revision{I_\tau}{\mathrm{I_\tau}}$}
  \end{minipage}
  \caption{%
    The mean settling velocity (normalized by the value for an
    isolated sphere in ambient fluid) in unbounded systems with forced
    background turbulence, plotted versus the relative turbulence
    intensity $I$.
    The filled circles %
    correspond to the parameter triplet ($\rho_p/\rho_f, \revision{Ga}{\mathrm{Ga}}, d_p/\eta$) as follows:
    {\color{black}$\bullet$}~($1.0004-1.04, 19-200, 12$)~\citep{fornari:16d};
    {\color{blue}$\bullet$}~($1.5, 120, 7$)~\citep{chouippe:15a};
    {\color{red}$\bullet$}~($1.5, 180, 7-9$)~\citep{chouippe:18};
    all aforementioned cases feature a global solid volume fraction
    $\Phi_s=0.005$ and background flow Reynolds number
    $Re_\lambda\sim\mathcal{O}(100)$. 
    The lines indicate the non-linear drag model of \citep{homann:13},
    evaluated for:
    {\color{black}\solid}~$\revision{Ga}{\mathrm{Ga}}=20$;
    {\color{black}\dashed}~$\revision{Ga}{\mathrm{Ga}}=180$.
    The data points marked by crosses are taken at the centerline of 
    vertical turbulent channel flow (cf.\
    figure~\ref{fig:PRR_vertChan_solidVolFrac_relvel}$b$ below, and
    the color-coding therein); in the channel flow cases the relative
    turbulence intensity is defined with the friction velocity
    $u_\tau$ as $\revision{I}{\mathrm{I}}_\tau=u_\tau/W_s$.
  }\label{fig:PRR_settling-and-turbulence}
\end{figure}

The analysis proposed by \citet{fornari:16a} indicates that large particles at low density ratios experience significant non-stationary effects. In a first approximation, one could model the influence of turbulence by adding a Gaussian perturbation with the same standard deviation as the forced turbulence to the velocity seen by the particles.
This would lead to the so-called nonlinear drag effect
\cite{homann:13,mei:97,bagchi:03}, which partially explains the trend
of the settling velocity reduction as shown in figure
\ref{fig:PRR_settling-and-turbulence}. 
Note that this model does neither account for the modification of the flow by the particles, nor for the particle-particle interaction through wake-attractions, nor for crossing trajectories.
It is indeed known that the two-way coupled background flow does not feature Gaussian like fluctuations in the vertical direction \cite{chouippe:18}, and as the solid volume fraction increases, turbulence has been shown to have less influence compared to the equivalent quiescent configuration. This last point is most probably to be attributed to the fact that drafting-kissing-tumbling remains relatively frequent at larger solid volume fraction \citep{fornari:18a}.
The relevant time-scales are also difficult to specify through a unique parameter (e.g. Stokes number based on the Kolmogorov timescale $\revision{St_\eta}{\mathrm{St_\eta}}=\tau_p/\tau_\eta$), first because particles are most probably affected by eddies larger than $\eta$ due to their size, and second since settling might decrease the time that particles spend in a given eddy altering the timescale actually perceived by the particles.
\section{PR-DNS of wall-bounded shear flows}
\label{sec:PRR-wall-bounded}
Wall-bounded shear flows are relevant to many technical and natural
systems, and they are therefore a popular laboratory for particulate
flow studies. Due to the presence of at least one inhomogeneous
spatial direction, the computational cost is significantly larger than
in unbounded set-ups, which explains the relatively recent advent of
PR-DNS in this category.
One clear advantage of wall-bounded flows is the natural occurrence of
turbulence at sufficiently large Reynolds number, which obviates the
need for ad hoc forcing.
It should be mentioned that the onset of laminar-turbulent
transition is greatly influenced by the presence of rigid particles 
\citep{matas:03}. This process which is the subject of
much ongoing research \citep{loisel:13,yu:13,agrawal:19}, however, reaches
beyond the scope of the present text.
Furthermore, in the following we will restrict our attention to plane
channel flow, since it constitutes the most widely studied
geometry. Other flow configurations have also been tackled by means of
PR-DNS, e.g.\ plane Couette flow
\citep{wang:17,alghalibi:18,rahmani:18}, 
circular pipe flow \citep{yu:13},
rectangular duct flow \citep{fornari:18,zade:19}. 
\subsection{%
  Vertical plane channel flow
}
\label{sec:PRR-wall-bounded-not-stratified}
In the present subsection we are considering configurations in which
gravity does not directly generate mean gradients in the particle
concentration. We are first focusing upon dilute or semi-dilute
suspensions with global \revision{solid}{particle} volume fraction up to
$\mathcal{O}(10^{-2})$, and then we will turn our attention to denser
systems.  
The questions which are most relevant in wall-bounded flows are
similar to those in unbounded flow (where do particles go, and what do
they do to the flow?) with the added complexity of spatial
inhomogeneity and anisotropy of the turbulent background flow.
\begin{figure}
  \begin{minipage}{2ex}
    \rotatebox{90}{%
      $\langle\phi\rangle/\Phi$
    }
  \end{minipage}
  \begin{minipage}{.45\linewidth}
    \centerline{$(a)$}
    \includegraphics[width=\linewidth]
    {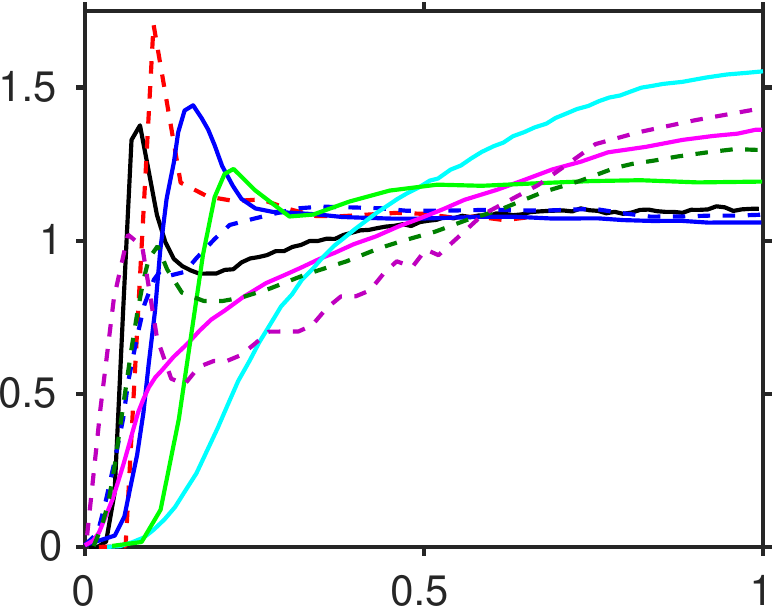}
    \hspace*{-.5\linewidth}\raisebox{.25\linewidth}{%
      \begin{minipage}{.45\linewidth}
        \includegraphics[width=\linewidth]
        {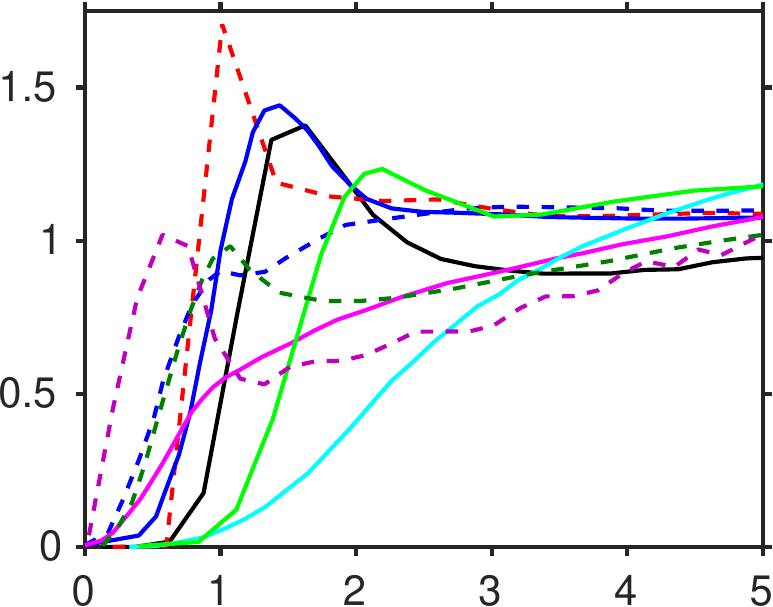}
        \hspace*{-5ex}\raisebox{2ex}{\small$y/d_p$}
      \end{minipage}
    }
    \\
    \centerline{$y/h$}
  \end{minipage}
  \hfill
  \begin{minipage}{.5\linewidth}
    \begin{minipage}{2ex}
      \rotatebox{90}{%
        $u_{rel}d_p/\nu$
      }
    \end{minipage}
    \begin{minipage}{.9\linewidth}
      \centerline{$(b)$}
      \includegraphics[width=\linewidth]
      {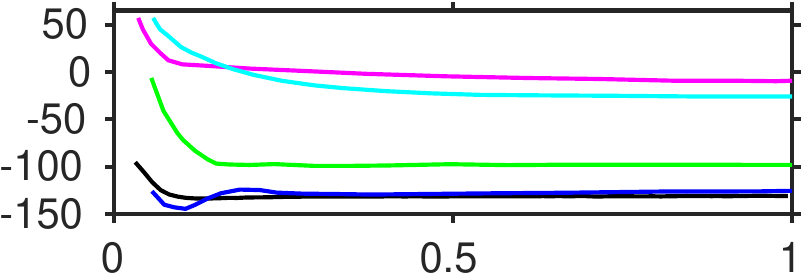}
      \\
      \centerline{$y/h$}
    \end{minipage}
    \\
    \begin{minipage}{2ex}
      \rotatebox{90}{%
        $u_{rel}^+$
      }
    \end{minipage}
    \begin{minipage}{.9\linewidth}
      \centerline{$(c)$}
      \includegraphics[width=\linewidth]
      {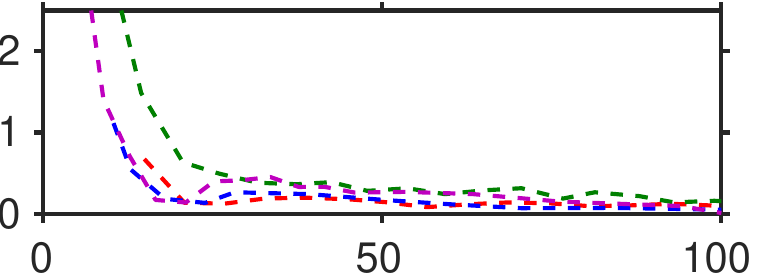}
      \\
      \centerline{$y^+$}
    \end{minipage}
  \end{minipage}
  \caption{%
    $(a)$
    Mean particle volume fraction $\langle\phi\rangle$ (normalized
    with the global value $\Phi$) in various plane channel flow PR-DNS
    without gravity effects in the wall-normal direction, plotted as
    function of wall-distance $y$, normalized by channel half-width
    $h$. 
    Solid lines correspond to cases where gravity acts on particles
    in the primary fluid flow direction, while dashed lines are for
    cases without gravitational effects. 
    The values for the parameter quadruplet ($\rho_p/\rho_f, Ga, d_p^+,
    \Phi_s$) are as follows: 
    {\color{blue}\dashed}~($1, 0, 21, 0.05$)~\citep{esteghamatian:21};
    {\color{red}\dashed}~($2.84, 0, 15,0.002$)~\citep{yu:16b};
    {\color{darkgreen}\dashed}~($10.4, 0, 18, 0.0084$)~\citep{yu:17}; 
    {\color{violet}\dashed}~($10, 0, 22,0.02$)~\citep{fornari:16c}; 
    {\color{magenta}\solid}~($2, 23, 19, 0.024$)~\citep{zhu:20}; 
    {\color{cyan}\solid}~($2, 40, 18, 0.024$)~\citep{zhu:20}; 
    {\color{green}\solid}~($2, 99.5, 21, 0.024$)~\citep{yu:21}; 
    {\color{black}\solid}~($2.2, 115.4, 11.3, 0.0042$)~\citep{villalba:12}; 
    {\color{blue}\solid}~($1.15, 126, 28, 0.05$)~\citep{esteghamatian:21}.
    The bulk Reynolds number measures $2700$--$2800$ over the entire
    data-set. 
    The inset shows the same data with the wall-distance scaled by the
    particle diameter. 
    $(b)$ Profiles of apparent velocity lag in the streamwise direction,
    $u_{rel}=\langle u_p\rangle-\langle u_f\rangle$, scaled with
    the viscous particle velocity, shown for cases with streamwise
    gravitational effects.  
    $(c)$ Same as $(b)$, but for cases without gravitational settling,
    scaled in wall units. 
  }\label{fig:PRR_vertChan_solidVolFrac_relvel}
\end{figure}
As a starting point let us discuss the mean relative velocity between
the two phases, which in the statistically stationary regime now
becomes a function of the wall-normal coordinate (henceforth denoted
as $y$, ranging from $0$ to $2h$, with $h$ the channel half-width). 
Figure~\ref{fig:PRR_vertChan_solidVolFrac_relvel}$(b,c)$ shows the
apparent relative velocity in the streamwise direction,
$u_{rel}=\langle u_p\rangle-\langle u_f\rangle$, for a number of
data-sets from PR-DNS of upward channel flow 
\citep{villalba:12,yu:16b,zhu:20,esteghamatian:21,yu:21}
conducted with neutrally-buoyant or dense particles at density ratios 
$\rho_p/\rho_f=\mathcal{O}(1)$, particle sizes
$d_p^+=\mathcal{O}(10)$ (where the $+$ superscript indicates wall
units) and bulk flow Reynolds numbers 2700 to 2800 (based upon the
bulk velocity $u_b$ and channel half width). It can be seen in the
figure that the relative velocity is essentially constant over most of
the bulk flow region, and that the magnitude of the bulk value
generally increases with increasing Galileo number from zero to
$\mathcal{O}(100)$. 
In order to gauge the effect of turbulence on mean settling, we have
included in figure~\ref{fig:PRR_settling-and-turbulence} 
the mean settling velocity on the channel centerline of those
simulations with finite Galileo number, normalized with the settling
speed of an isolated particle in ambient fluid (obtained from
equilibrium between drag and submerged weight, resorting to the
standard drag law \citep{clift:78}), 
plotted versus the relative turbulence intensity now defined as 
$I_\tau=u_\tau/u_g$
\revision{}{(where $u_\tau$ is the mean friction velocity).} 
As in the case of unbounded flow with homogeneous-isotropic forcing, a
significant decrease of the settling speed with increasing $I_\tau$
is observed, which again might be related to the non-linear drag
mechanism already discussed in
subsection~\ref{fig:PRR_settling-and-turbulence}.  
Please recall that the global \revision{solid}{particle} volume
fraction varies among the present channel flow data points (cf.\ the
caption of figure~\ref{fig:PRR_vertChan_solidVolFrac_relvel}), 
which is expected to contribute to the reduced settling velocity
through the Richardson-Zaki hindrance effect.  

Interestingly, all curves in
figure~\ref{fig:PRR_vertChan_solidVolFrac_relvel}$(b)$ exhibit an
upward trend when approaching the channel wall, indicating a
further reduction of the particle settling velocity in the near-wall
region. The same is also true for neutrally-buoyant particles and for
dense particles in the absence of gravity (for which both $Ga=0$), as
can be seen in figure~\ref{fig:PRR_vertChan_solidVolFrac_relvel}$(c)$.
In upward channel flow with heavy particles \citet{uhlmann:08a} has
attributed this effect to an increase in the quasi-steady drag force
when a particle is located at a wall distance of the order of its
diameter (cf.\ drag correlations for a sphere translating near a wall,
\citep{zeng:05}, and for a sphere translating in a linear,
wall-bounded shear flow \citep{lee:10a}).  
Other authors have shown that particles near the wall 
preferentially sample the high-speed regions in upward channel flow
with heavy particles \citep{zhu:20}, as well as in channel flow with
dense particles in the absence of gravity \citep{yu:16b}. 
In both cases this sampling bias is consistent with the observed
trend, which is reduced when instead computing the mean relative
velocity with the fluid velocity measured in the vicinity of the
particle.

The next quantity of interest is the mean particle concentration which
is shown in figure~\ref{fig:PRR_vertChan_solidVolFrac_relvel}$(a)$ for
the same parameter points as previously discussed. It can be seen that
the shape of the profiles varies strongly, ranging from parabola-like
(for dense particles at small but finite Galileo numbers) to nearly
flat in the bulk (at large Galileo number, but also for some
zero-Galileo cases) with marked local (or global) peaks in the
near-wall region. The inset of
figure~\ref{fig:PRR_vertChan_solidVolFrac_relvel}$(a)$ shows that the
near-wall peak occurs at wall-distances between $0.5d_p$ (particle in
contact with the wall) to approximately $2d_p$. 

A number of mechanisms have been proposed in the
literature to account for the mean particle distribution:
(i) mean wall-normal (lift) force \citep{uhlmann:08a};
(ii) turbophoresis \citep{uhlmann:08a};
(iii) collisions \citep{fornari:16c}. 
The mean lift effect, which can be decomposed into contributions due
to shear, translation and rotation, can be gauged with the aid of the
quasi-steady force correlations given by \citep{lee:10a}.
\citet{esteghamatian:21} conclude that the rotational contribution is
important in establishing a near-wall peak as it drives particles on
the wall-facing edge of a cluster towards the wall, thereby countering
the action of shear-induced lift which tends to drive particles in the
opposite direction. \citet{uhlmann:08a} argued that turbophoresis --
the transport of particles down gradients of turbulence intensity
\citep{caporaloni:75,reeks:83} --
is the main actor which drives particles towards the wall against 
shear-induced wall repulsion.
\citet{fornari:16c} argued that in the absence of gravity and for dense
particles, collisions are the most effective mechanisms in driving
particles towards the channel center.
A more complete quantification of the particle distribution has been
performed by means of various criteria, such as box-counting
\citep{uhlmann:08a}, 
nearest-neighbor statistics \citep{kajishima:01},
radial distribution functions \citep{fornari:16c}, 
and Vorono\"i tesselation \citep{villalba:12,esteghamatian:21}.
Thereby, extensive information on the macro and micro-structure of the
disperse phase has been obtained, such as evidence of drafting at
sufficiently large Galileo number
\citep{villalba:12,esteghamatian:21}. However, many fundamental 
questions still remain. It is for example not clear why the (small,
heavy) particles in \citep{villalba:12} exhibit a slightly more
ordered structure than a fully random draw, while those in
\citep{esteghamatian:21} (which are larger) show a clear degree of 
clustering.  
From this short overview it can be concluded that a full understanding
of the dynamics leading to the spatial distribution of the disperse 
phase over the entire parameter space in wall-bounded flows has not
yet been achieved. 

Now let us briefly discuss particulate channel flow with
neutrally-buoyant particles in the dense regime with a global
\revision{solid}{particle} volume fraction $\mathcal{O}(10^{-1})$.
Here the contribution from the particle stress becomes comparable to
the fluid Reynolds stress in the momentum balance, and the turbulence 
structure is signifcantly modified \citep{picano:15}. 
Several authors \citep{lashgari:14,picano:15,costa:16,costa:18} have
observed that particles in this case tend to form a relatively dense
layer adjacent to the wall, where they also exhibit a significant
value of the apparent slip velocity (cf.\ the discussion on the
scaling of the velocity slip in \citep{costa:18}).
Separately modelling the stresses in the particle layer and
the bulk flow region (where an effective viscosity can be used) then
leads to scaling laws for the mean velocity and the wall friction as
functions of \revision{solid}{particle} volume fraction and particle size
\citep{costa:16}.
\citet{lashgari:17} have shown that this reasoning still mostly holds
when the disperse phase consists of a binary mixture of small and
large particles ($d_p^+=18$ and $24$), while the main effect of
bi-dispersity lies in the details of the near-wall particle layer. 

The literature on wall-bounded turbulence in flows laden with
non-spherical particles is rather scarce. Only spheroids of various
aspect ratio, spanning both oblate and prolate spheroids, have been
considered either in plane channel flows
\citep{ardekani:17b,eshghinejadfard:18,zhu:18,eshginejadfard:19,ardekani:19,zhu:20}
or in pipe flows
\citep{gupta:18}. 
To the best of the authors'
knowledge, there is no account in the literature of any study
involving finite-size cylinders or polyhedrons. The existing body of
knowledge on spheroids spans neutrally buoyant
\citep{ardekani:17b,eshghinejadfard:18,zhu:18,gupta:18,ardekani:19}, 
non-neutrally buoyant without gravitational effect
\citep{eshginejadfard:19}, 
and non-neutrally buoyant with
gravitational effect
\citep{zhu:20}.
The overall picture of the effect of shape via the spheroid aspect
ratio when compared to spheres includes the measurable reduction of
flow drag and turbulent fluctuations, the stronger migration of
spheroids to the center of the channel, the depletion of spheroids
in the region close to the wall, the lower angular velocity of
spheroids, and the tendency of spheroids to orient parallel to the
wall. It is also observed that the more the spheroid aspect ratio
deviates from unity
\citep{zhu:20} 
or the larger the particle to fluid density ratio is
\citep{eshginejadfard:19} 
the more pronounced
the aforementioned flow features are. 
\def\asolidthick{\protect\rule[2pt]{10.pt}{1pt}}
\def\asolidshort{\protect\rule[2pt]{3.pt}{1pt}}
\def\adashed{\asolidshort$\,$\asolidshort$\,$\asolidshort}
\def\achndot{\asolidshort$\,\cdot\,$\asolidshort}
\def\asolidthickdotted{\asolidthick$\hspace{-1ex}
                      \bullet\hspace{-1ex}$\asolidthick}
\def\asolidthicksquared{\raisebox{3pt}{\asolidthick}\hspace{-2ex}
                      \asolidsquare\hspace{-1ex}\asolidthick}
\def\asolidthickopensquare{\raisebox{0.5pt}{\asolidthick}\hspace{-2ex}
                      \aopensquare\hspace{-1ex}\raisebox{0.5pt}{\asolidthick}}
\def\asolidthickopencircle{\raisebox{1.5pt}{\asolidthick}\hspace{-2ex}
                      \aopencircle\hspace{-1ex}\raisebox{1.5pt}{\asolidthick}}                         
\def\asolidthickbullet{\asolidthick\hspace{-1ex}
                      \bullet\hspace{-1ex}\asolidthick}
\def\asolidthicktriup{\asolidthick\hspace{-1ex}
                      \blacktriangle\hspace{-1ex}\asolidthick}
\def\asolidthicksquare{\asolidthick\hspace{-1ex}
                      \mathsmaller{\blacksquare}\hspace{-1ex}\asolidthick}

\usetikzlibrary{shapes.geometric,shapes.symbols}
\newcommand{\tikzsymbol}[2][circle]{\tikz[baseline=-0.5ex]\node[inner
sep=2pt,shape=#1,draw,#2]{};}%
\newcommand{\amanbullet}[2]{%
  \tikzsymbol[circle]{#1,fill=#2}%
  }
\newcommand{\amanellipse}[2]{%
  \tikzsymbol[ellipse]{#1,fill=#2,minimum width=8pt}%
  }  
\newcommand{\amandiamond}[2]{%
  \tikzsymbol[diamond]{#1,fill=#2,scale=0.9}%
  }
\newcommand{\amanstar}[3]{%
  \tikzsymbol[star]{#1,fill=#2,star points=#3,star point ratio=2.25,scale=0.65}%
  }
\newcommand{\amansquare}[2]{%
  \tikzsymbol[rectangle]{#1,fill=#2,scale=1.3}%
  }
\newcommand{\amantriright}[2]{%
  \tikzsymbol[isosceles triangle,isosceles triangle apex angle=60,rotate=0,anchor=center]%
  {#1,fill=#2,scale=0.8}%
  }
\newcommand{\amantriup}[2]{%
  \tikzsymbol[isosceles triangle,isosceles triangle apex angle=60,rotate=90,anchor=center]%
  {#1,fill=#2,scale=0.8}%
  }
\newcommand{\amantrileft}[2]{%
  \tikzsymbol[isosceles triangle,isosceles triangle apex angle=60,rotate=180,anchor=center]%
  {#1,fill=#2,scale=0.8}%
  }
\newcommand{\amantridown}[2]{%
  \tikzsymbol[isosceles triangle,isosceles triangle apex angle=60,rotate=270,anchor=center]%
  {#1,fill=#2,scale=0.8}%
  }  

\newcommand{\across}{$\mathlarger{\mathlarger{\times}}$}
\definecolor{amanred}{rgb}{0.8,0,0}
\definecolor{amanblue}{rgb}{0,0,0.7}
\definecolor{myblue}{rgb}{0,0.4,0.8}
\definecolor{KITred}{rgb}{0.63,0.12,0.1}
\definecolor{myred}{rgb}{0.63,0.12,0.1}
\definecolor{mygreen}{rgb}{0,0.6,0.51}
\definecolor{mygray}{rgb}{0.63,0.63,0.63}
\definecolor{myorange}{rgb}{0.9,0.55,0.0}
\definecolor{myturk}{rgb}{0.13,0.42,0.52}
\definecolor{mypurple}{rgb}{0.66,0.42,0.61}

\newcommand\prdns{PR-DNS}
\newcommand\shields{\ensuremath{\Theta}}
\newcommand\utau{\ensuremath{u_\tau}}
\newcommand\dratio{\ensuremath{\rho_p/\rho_f}}
\newcommand\grav{\ensuremath{\mathbf{g}}}
\newcommand\Ret{\ensuremath{Re_\tau}}
\newcommand\Reb{\ensuremath{Re_b}}
\newcommand\hfluid{\ensuremath{H_f}}
\subsection{Horizontal plane channel flow}
\label{sec:PRR-wall-bounded-hchan}
Here we are interested in the configuration where gravity is directed in the wall-normal direction, such that heavy particles will tend to settle towards the bottom wall. As a direct consequence the mean particle distribution is inhomogeneous, and the details will again depend on their complex interplay with the turbulent shear flow and its coherent structures. This system is particularly relevant to open water bodies (such as rivers and streams), whence the channel is bounded at the top by a free surface.      
\subsubsection{Focusing of heavy particles in low-speed streaks}
\label{sec:PRR-wall-bounded-hchan-settling}
Various experimental and numerical studies have shown that
buffer layer coherent structures play an
important role in the dynamics of particle motion.
Heavy particles, depending on their size, their density and flow Reynolds number,
get entrained to the outer flow by the action of the coherent structures.
While resettling, the particles are observed to migrate and accumulate into
the low-speed fluid regions forming streamwise-aligned streaky particle clusters
\citep{kaftori:1995a,nino:1996,Shao2012,Kidanemariam2013}.
In one of the earliest \prdns\ of the configuration,
\citet{Kidanemariam2013} have shown via conditional averaging
that particle lateral motion is strongly correlated with the
rotation-sense of a nearby quasi-streamwise vortices. The statistical
analysis explains the fact that migration of heavy particles to
low-speed streaks is due to the action of the counter-rotating
streamwise vortices. It is argued that this phenomenon is the main
reason for the observed lower average streamwise particle velocity
than the mean fluid velocity in this configuration.
As a proof of concept,
\citep{pestana:19}
have recently revisited the 
above-discussed particle migration into low-speed streaks but by considering
equilibrium solutions which feature exact coherent structures (streaks and vortices) to
serve as a surrogate of the near-wall turbulent structures. They have shown that
particles indeed laterally migrate to the low-speed streaks,
demonstrating robustness of the mechanism.
\subsubsection{Sediment transport}
\label{sec:PRR-wall-bounded-sediment-featureless}

Sediment transport is a dense particulate flow problem which 
involves the erosion and  deposition of sediment grains as a net result of
driving hydrodynamic forces and resisting gravity and inter-particle
\revision{}{collision} forces.
Depending on the relative dominance of these forces, sediment grains
in a given stream may be stationary or are set in motion by the flow.
The onset of sediment
motion and subsequent transport is generally
believed to be controlled by the Shields number
\revision{}{\citep{Shields1936}:}
\begin{equation}\label{eq:PRR_def-shields-general-case}
  \shields =
  \frac{\utau^2}{\left(\dratio-1\right)|\grav|d_p}
  = \left(\frac{\utau}{u_g}\right)^2
  = \left(\frac{d_p^+}{Ga}\right)^2
    \,,
\end{equation}
\revision{}{where}
$|\grav|$ and $d_p$ are the magnitude of acceleration due to gravity and
the particle diameter respectively.
\dratio\ is the particle to fluid density ratio while
${u_g = \sqrt{(\dratio -1) g d_p}}$ is the gravitational velocity scale.
\revision{}{\shields\ is a measure of the ratio of 
the total shear force acting on a single grain of the sediment bed to the
apparent weight of the particle.}
For a given flow condition, there is a critical Shields number value $\shields_c$
below which the shear force induced by the flow is unable to dislodge and entrain 
particles from the bed and no sediment erosion is observed.
$\shields_c$ is known to depend on a number of flow and sediment bed parameters
including the Reynolds number of the flow,
the constituent sediment grain packing, grain shape, polydispersity,
inter-particle contact, etc.
At supercritical values of the Shields number ($\shields > \shields_c$),
larger grains are typically transported close to the bed as 
`bedload'. In bedload transport mode, particles experience regular 
contact with the bed and move by rolling, sliding or in a series 
of hopping motions of short duration called saltation. On the other hand,
in the case of fine grains or flow with relatively large flow velocities
($\shields \gg \shields_c$), particles are generally 
transported as `suspended load' and remain in suspension for longer times
by processes of advection and turbulent diffusion \citep{Garcia2008}.  In the bedload regime,
the issuing sediment flux
is often expressed as a function of the Shields number
(or the excess Shields number $\shields - \shields_c$) and
various  \mbox{(semi-)empirical} predictive models have been proposed
\citep[see e.g.][]{Meyer-Peter1948} \revision{}{and are widely used in
engineering practice.}

Understanding the physics of sediment transport at the grain scale is imperative
to critically assess and improve the various existing predictive models.
This is where \prdns\ comes in, which has been
increasingly exploited in recent
years to simulate subqueous sediment transport phenomena, both in the
laminar and turbulent regimes. Due to the ability of the method to
resolve the fluid-particle interaction at the grain scale, \prdns\ has
proven to be instrumental in complementing existing experimental and
theoretical works by allowing access to relevant quantities.
In the following, we \revision{focus on}{briefly review} 
\prdns\ based studies of sediment transport
induced by a shearing flow, focusing exclusively
on the subaqueous regime, i.e.\ $\dratio=\mathcal{O}(1)$.

\begin{figure}[t]
  \centering
  \begin{minipage}{2ex}
    \rotatebox{90}{\hspace{4ex}$\langle q_p\rangle /q_{v}$}
  \end{minipage}
  \begin{minipage}{0.3\textwidth}
    \centerline{(\textit{a})}
    \includegraphics[width=\linewidth]{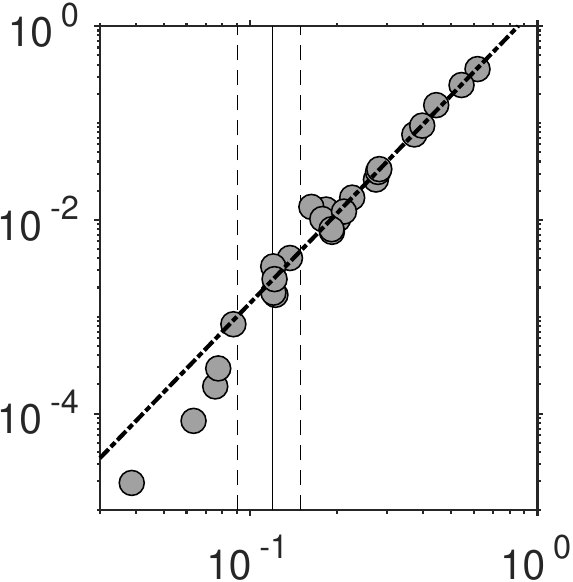}
    \centerline{$\shields_{Pois}$ }
  \end{minipage}
  \hfill
   \begin{minipage}{2ex}
    \rotatebox{90}{\hspace{4ex}$\langle q_p\rangle /q_{i}$}
  \end{minipage}
  \begin{minipage}{0.3\textwidth}
    \centerline{(\textit{b})}
    \includegraphics[width=\linewidth]{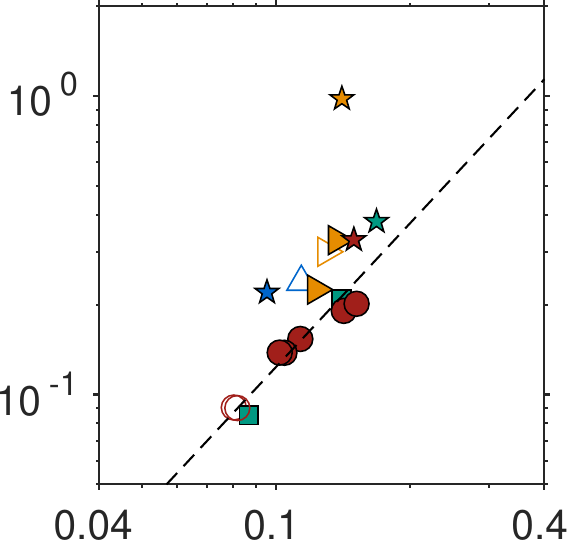}
    \centerline{$\shields - \shields_c$ }
  \end{minipage}         
  \hfill
   \begin{minipage}{2ex}
    \rotatebox{90}{\hspace{4ex}$k_0/d_p$}
  \end{minipage}
  \begin{minipage}{0.27\textwidth}
    \centerline{(\textit{c})}
    \includegraphics[width=\linewidth]{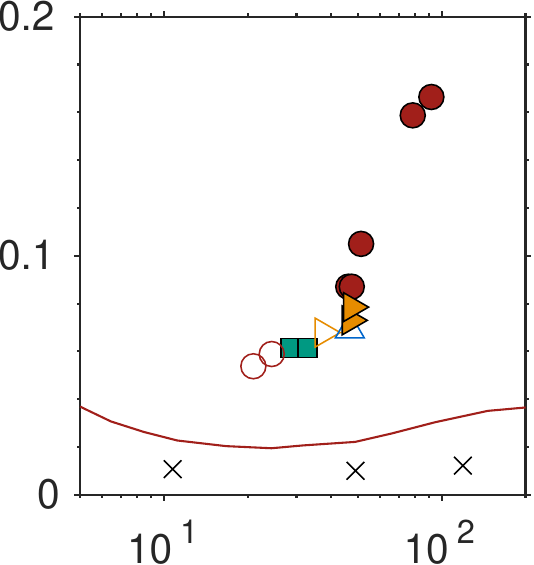}
    \centerline{$d_p^+$, $H_D^+$}
  \end{minipage}
  \caption{%
   (\textit{a}) \prdns\ of  a sediment bed sheared by laminar channel flow:
   Mean volumetric particle flow rate $\langle q_p \rangle$, non-dimensionalized by the viscous
   scaling ${q_v = Ga^2\nu}$, as a function of the Shields number defined based on
   the laminar plane Poiseuille flow over a smooth wall, viz
   ${\shields_{Pois} = 6\Reb/Ga^2(d_p/\hfluid)^2}$.
   \protect\amanbullet{black}{gray},
   $\Reb=110$--$500$, $Ga = 6$--$15$, $\hfluid/d_p=7.5$--$20$, $\dratio = 2.5$
   \citep{Kidanemariam2014b}.
   The vertical solid/dashed lines indicate the critical Shields number for
   particle erosion $\shields_{Pois}^c = 0.12 \pm 0.03$ \citep{Ouriemi2007}.
   The chain-dotted line corresponds to a power law fit proportional to $\shields_{Pois}^3$.
   (\textit{b})
   \prdns\ of sediment bed erosion in turbulent channel flow.
   Volumetric particle flow rate $\langle q_p\rangle$ (normalized by the
   inertial scale $q_i = u_g d_p$)
   as a function of the excess Shields number $\shields-\shields_c$.
   The critical Shields number $\shields_c$ is obtained using the empirical law
   proposed by \citet{soulsby:97}.
   \amanbullet{black}{myred} ripple-featuring, \amanbullet{myred}{white} ridge-featuring,
    $Ga = 28$, $\hfluid/d_p=25$, $\dratio = 2.5$, $d_p^+=9$--$12$;
    \amansquare{black}{mygreen} ripple-featuring,
    $Ga = 17$--$20$, $\hfluid/d_p=50$, $\dratio = 2.5$, $d_p^+=7$;
    \amantriup{myblue}{white} featureless,
    $Ga = 105$, $\hfluid/d_p=8$, $\dratio = 2.5$, $d_p^+=39$;
    \amantriright{black}{myorange} ripple-featuring,
    \amantriright{myorange}{white} featureless,
    $Ga = 44$--$57$, $\hfluid/d_p=12$--$16$, $\dratio = 2.5$, $d_p^+=17$--$23$
    \citep{Kidanemariam2017,Scherer2020}.
    The star symbols correspond to the \prdns\ of ellipsoidal particles
    (\amanstar{black}{myred}{5}, Zingg ellipsoid;
     \amanstar{black}{mygreen}{5}, prolate;
     \amanstar{black}{myorange}{5}, oblate;
     \amanstar{black}{myblue}{5}, sphere)
    $Ga = 45$, $\hfluid/d_p=16$--$18$, $\dratio = 2.55$, $d_p^+=16$--$20$
    \citep{Jain2021}.
    The dashed line is
    the \citet{Wong2006} version of the \citet{Meyer-Peter1948}
    formula for turbulent flows:
   $q_p/q_i=4.93\,(\shields-\shields_c)^{1.6}$.
   (\textit{c})
   The hydrodynamic roughness length $k_0$, obtained by fitting the relation
   $\langle u_f\rangle^+ = 1/\kappa\log((y-y_0)/k_0)$ to the fluid mean velocity profiles
   of the corresponding \prdns\ cases in
   \ref{fig:PRR_particle-flow-rate-vs-shields}\textit{b}, as a function
   of the inner-scaled sediment bed roughness height $H_D^+$
   \revision{from maximum to minimum}{($H_D$ is the crest-to-trough height
   of the extracted two-dimensional sediment bed interface)}.
   The cross symbols (\across) represent \revision{}{values of $k_0$ as a function
   of $d_p^+$ of a} single-phase flow over fixed spheres
   arranged on a square lattice \citep{Chan-Braun2011}.
   The solid line is the model prediction \revision{}{$k_0$ as a function of $d_p^+$}
   by \citet{DuranVinent2019}.
   }
   \label{fig:PRR_particle-flow-rate-vs-shields} 
\end{figure}

Let us \revision{mention}{remark} that performing \prdns\ of sediment transport,
even in today's computing power, is immensely  expensive
\revision{}{(see example the challenges discussed in section
\ref{sec:PRR-wall-bounded-sediment-patterns})}.
The ability to faithfully resolve the underlying fundamental mechanisms
without modelling comes thus at the cost of simplifying the flow configuration,
both in terms of the parameter space and system size.
Another aspect of importance is the inter-particle interaction
which contributes significantly to the dynamics of the system.
In order to realistically account for the collision
process between the submerged grains, a discrete element model (DEM)
based on either the soft-sphere or hard-sphere approach is usually
incorporated to the \prdns\ solution strategy (see details in
chapter~7). 
Some of the earliest contributions have considered the erosion of a 
sediment bed made up of mono-dispersed spherical particles,
induced by shearing laminar flow
\citep{Derksen2011,Kidanemariam2014b}.
These studies were motivated, in part, by the availability of
detailed experimental data to assess the level of fidelity
achieved by the numerical approach.
\citet{Derksen2011} performed \prdns\ of sediment erosion in a
Couette flow configuration
while \citet{Kidanemariam2014b} simulated
sediment transport in a laminar Poiseuille
covering a wide range of Shields number values. These studies have
successfully reproduced experimental observations regarding
the threshold for sediment erosion $\shields_c$ \citep{Ouriemi2007}
and the scaling of the sediment particle flux
for $\shields>\shields_c$ \citep{Aussillous2013}.
Figure \ref{fig:PRR_particle-flow-rate-vs-shields}\textit{a} shows the 
particle flow rate as a function of the Shields number obtained from the
\prdns\ by \citet{Kidanemariam2014b}, recovering the well-known 
cubic variation of the particle flow rate with the
Shields number for $\shields>\shields_c$.
Furthermore, the direct access to the hydrodynamic and collision forces
acting upon individual particles in \prdns\ has
enabled the investigation of the rheological proprieties of the mobile sediment bed
and the contribution of fluid and particle stresses to the momentum
transfer between the two phases
\citep{Kidanemariam2015,Vowinckel2021}.

In the turbulent regime, most of available \prdns\ of sediment transport phenomena
are at moderate Reynolds number values (friction velocity based Reynolds number
$\Ret=\mathcal{O}(100)$) and have mainly considered a single or a few
particles to investigate the role of the near-wall turbulent structures on the
entrainment mechanism and subsequent trajectories 
\citep{Munjiza2000,chan-braun:12,Vowinckel2016,Jain2017,Yousefi2020}. There are
also studies which consider a few layer of mobile particles to study
the collective sediment transport and its modulation of the turbulent flow
\citep{Ji2013,Ji2014,Vowinckel2014,Derksen2015,Vowinckel2017a,Vowinckel2017b}.
\citet{Derksen2015}
simulated granular bed erosion in a mildly turbulent channel flow
and looked at effect of Shields number on the mobility of particles and
addressed the role of turbulent fluctuations on particle erosion.
\citet{Vowinckel2014} considered the transport of sediment particles over
a rough bottom
and investigated the particle transport patterns
and their influence on the bulk statistics of the flow. Extending
\revision{the}{their} work,
\citet{Vowinckel2017a,Vowinckel2017b} analysed the momentum balance
in a similar setup. Based on the double-average methodology, the latter authors
assessed the turbulent and form-induced momentum fluxes as well as boundary stresses
contributions to the total budget and the role of mobile particles in the distribution
of these stresses.
Let us mention that the \prdns\ of turbulent
flow over a thick layer of mobile sediment by \citet{Kidanemariam2014} and
subsequent publications are discussed in the context of pattern
formation in the next section \ref{sec:PRR-wall-bounded-sediment-patterns}.

Most available \prdns\ of sediment transport have considered spherical
particles. However, it is known that particle shape and orientation
are important parameters \revision{in sediment transport}{of the flow configuration}.
There are recent \prdns-based studies
\revision{which}{that address this issue and} consider non-spherical sediment grains
\citep{Fukuoka2014,Fukuda2019,Zhang2020,Jain2020,Jain2021}.
For instance, \citet{Jain2021} have considered 
ellipsoidal particles to investigate the effect of particle shape on
their trajectories, pickup, erosion and deposition rates and the
resulting collective effect
(cf.\ figure \ref{fig:PRR_particle-flow-rate-vs-shields}\textit{b})
These studies highlight the promising prospect of \prdns\ to tackle complex
sediment transport phenomena in the near future. 
\subsubsection{Sediment pattern formation}
\label{sec:PRR-wall-bounded-sediment-patterns}
\begin{figure}[t]
 \centering
 \begin{minipage}{\textwidth}
   (\textit{a})\hfill\mbox{}\\
   \includegraphics[width=\textwidth]
    {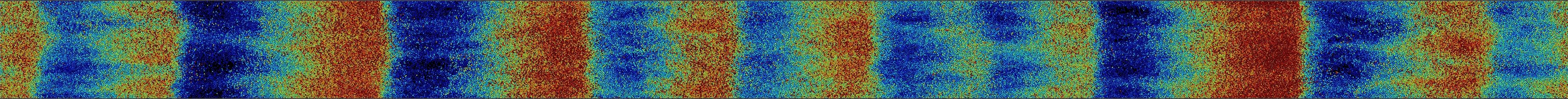}
    \centerline{$L_x \approx 48\hfluid \approx 1200d_p$;
                resolved particles $\approx 1.1\times 10^{6}$ }             %
   \mbox{}\\ 
   \begin{minipage}{4ex}       
     (\textit{b})
   \end{minipage}
   \begin{minipage}{2ex}       
     \rotatebox{90}{\hspace{-2ex}$y/d_p$ \hspace{12ex} $z/d_p$}
   \end{minipage}
   \begin{minipage}{0.6\textwidth}
     \includegraphics[width=\linewidth] %
         {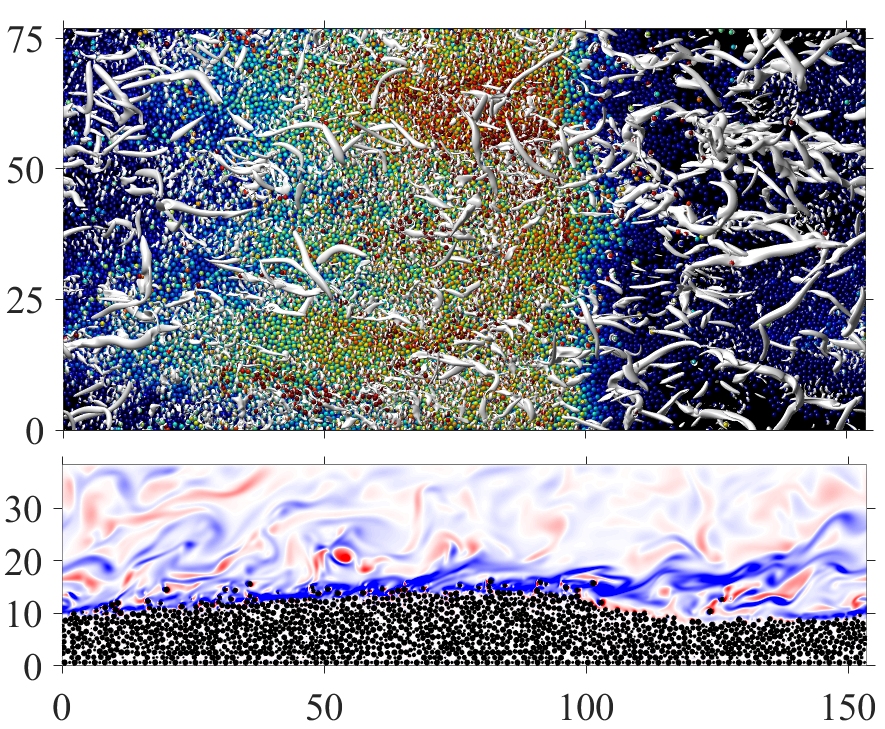}\\[-5pt]
         \centerline{$x/d_p$}
   \end{minipage}
   \hspace{2ex} 
   \begin{minipage}{0.27\textwidth}
     \centering
     Particle location\\ $y_p/d_p$\\
     \includegraphics[width=0.9\linewidth]{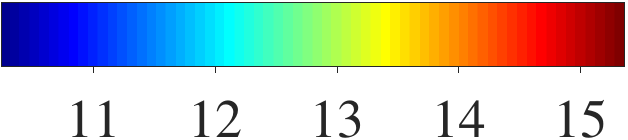}\\[30pt]
     Spanwise vorticity\\  $\omega_f^+ = \omega_f \nu/\utau^2$\\
     \includegraphics[width=\linewidth]{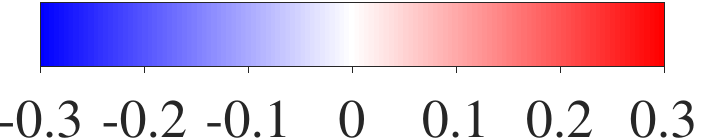}
   \end{minipage}
 \end{minipage}
 \caption{Instantaneous visualisations of the flow field and particle
   positions in a turbulent channel flow over an evolving sediment bed.
   $\Reb = 3010$; $Ga = 28.4$; $\hfluid/d_p=25$; $\dratio = 2.5$;
   $\Ret = 240$--$310$; $d_p^+ = 9$--$12$
   \revision{}{\citep{Kidanemariam2017,Kidanemariam2021a:submitted}:}
  (\textit{a})
  Top view of the sediment bed which is represented by
  approximately 1.1~million spherical particles,
  accommodating a series of self-forming transverse ripples
  \revision{which}{that} have formed after approximately \revision{$200$}{$400$}
  bulk time units after initial particle release
  \revision{(taken from \citet{Kidanemariam2017})}{}.  Sediment
  grains are coloured according to their wall-normal location $y_p$.
  Flow direction is from left to right.
 (\textit{b}) 
  Snapshot of the near-wall turbulent vortical structures over a single
  sediment ripple unit (which was forced to steadily evolve in a box size
  of $L_x \approx 150d_p$) along with the corresponding spanwise vorticity shown on a
  select wall-perpendicular plane
  \revision{(taken from \citep{Kidanemariam2021a:submitted}}{}.
  The strong modulation of the
  turbulent flow by the evolving sediment bed is highlighted by the increased
  turbulence intensity downstream of the crest of the ripple.}
   \label{fig:PRR_ripple-visualization}
\end{figure}
Probably the most notable achievement of \prdns\ in the context of
sediment transport has been the ability to simulate the phenomenon of
sediment pattern formation
\citep{Kidanemariam2014,Kidanemariam2017,Mazzuoli2019,
  Scherer2020,Jain2021,Kidanemariam2021a:submitted,Scherer2021a:submitted}.
The process of local erosion of particles from an erodible sediment
bed and their deposition at certain preferential locations often leads
to the formation of sediment patterns or bedforms. Sediment patterns
are observed at a wide range of spatial and temporal scales in a
variety of environments such as deserts, river and marine flows as
well as various technical applications involving shear flow over an
erodible sediment bed. It is highly desirable to accurately predict
the formation processes of bedforms as they have important
implications for instance in
waterway management, planning and maintenance of various hydraulic
structures and geophysics.
Capturing sediment pattern formation and evolution via
\prdns\ is immensely challenging due to the much larger spatial
and temporal scales of the patterns (compared to the particle scale)
which need to be resolved. In order to realistically capture the patterns,
large computational domains involving up to $\mathcal{O}(10^{10})$
grid nodes are required. Moreover, a thick erodible sediment bed composed of
at  least  $\mathcal{O}(10^5)$--$\mathcal{O}(10^7)$ particles
must be considered in order to accommodate patterns.
\revision{}{At the same time, the ratio between the particle size and the
grid cell width should be adequately large (typically $d_p/\Delta x = \mathcal{O}(10)$)
in order to resolve the near-field flow
evolving around individual particles.}
The evolution time scales of the
patterns is another constraint which requires simulation integration times of
at least $\mathcal{O}(10^3)$ bulk time units.
\revision{These}{The above}
factors pose severe computational challenge
and current \prdns\ simulations are so far mainly restricted to the
initial patterns formation and evolution  stages only.

Customarily, the flow over an erodible bed has been theoretically
treated as a hydro-morphodynamic instability problem and is usually
tackled by linear stability analysis.  The hydrodynamics is coupled to
the morphodynamic evolution of the bed by an algebraic expression for
the particle flux as a function of the Shields number and prediction
is made regarding the stability of the sediment bed, the controlling
parameters of the instability as well as the initial pattern wavelength
\citep{Charru2013a,Andreotti2013}.
However, the prediction is mostly unsatisfactory which can be
linked to, among others, the deficiency of the adopted algebraic expressions
for the particle flux. Furthermore, the subsequent evolution of the bedforms
and the interaction with the background turbulent flow is not captured by linear analysis.
\prdns\ is being increasingly utilised to tackle some of these outstanding questions.
\citet{Kidanemariam2014,Kidanemariam2017} and later \citet{Scherer2020}
carried out \prdns\ of the formation of sediment patterns in a
channel flow configuration considering a very large number of mobile sediment grains
focusing on the initial wavelength selection mechanism, identifying the lower threshold
of unstable wavelength and its scaling as well as aspects of the steady non-linear
evolution, ripple migration velocity and ripple morphological characterstics.
\citet{Kidanemariam2021a:submitted} have further analysed the modulation of the turbulent
flow by the evolving sediment bed (cf.\ {figure \ref{fig:PRR_ripple-visualization}}
which  clearly show the complex feedback between the  flow and the evolving sediment bed).
In particular, an evaluation of the local momentum budget was carried
out to scrutinise the spatial variation of the boundary shear stress
variation. The accurate determination of this quantity, which is one
of the main ingredients in models of sediment bed morphology, has been
an outstanding issue. The analysis has shown that the shear stress is
maximum at a location upstream of the crests exhibiting a positive
phase shift with respect to the ripple geometry while the
spatially-resolved particle flux lags the shear stress.  The phase lag
of the particle flux with respect to the shear stress is a consequence
of sediment inertia, i.e.\ a relaxation of the sediment flux to
changes in the shear stress \citep{Charru2013a}.

Central in many sediment transport models are the algebraic expressions
which relate the mean particle flux $\langle q_p \rangle$ with the  Shields number.
\prdns\ data has been recently used to assess the validity of such models
\citep{Kidanemariam2014,Kidanemariam2017}.
Figure \ref{fig:PRR_particle-flow-rate-vs-shields}\textit{b} shows \prdns\
results of the variation of $\langle q_p \rangle$ as a function of the excess
Shields number $\shields - \shields_c$ from different cases of a sheared sediment
bed with and without bedforms. It can be seen that the mean particle flow rate
is well predicted by power-law type formulas such as that by \citet{Wong2006}.
Nevertheless, as has been recently reported by \citep{Kidanemariam2021a:submitted},
such power laws are unable to predict the local relationship between the particle
flux and the shear stress as a consequence of sediment flux relaxation.

Turning to the the feedback of the evolving rough sediment bed on the turbulent flow,
one way of quantifying it is to scrutinise the influence on the vertical variation of the
mean fluid velocity profile $\langle u_f \rangle$.
Commonly, $\langle u_f \rangle$ in the logarithmic layer is expressed
through the definition of a hydrodynamic roughness height $k_0$ as
$\langle u_f \rangle ^+ = 1/\kappa \log((y-y_0)/k_0)$ where $\kappa=0.41$ is the von
K\'{a}rm\'{a}n coefficient and $y_0$ is some reference  origin \citep{Jimenez2004}.
The majority of the available sediment morphodynamic models strongly depend on the
roughness parameter $k_0$ \citep{Charru2013a},
and make use of classical correlations of $k_0$ versus particle Reynolds number $d_p^+$
which are obtained, for example, from flows over stationary roughness elements
(see e.g \citep{DuranVinent2019}). However, as is demonstrated in
{figure \ref{fig:PRR_particle-flow-rate-vs-shields}\textit{c}},
the motion of the sediment bed and the evolving ripples substantially increase
the value of $k_0$ for essentially the same value of $d_p^+$. For comparison, $k_0$
values obtained from \prdns\ of the flow over stationary roughness
\citep{Chan-Braun2011} are also shown in the figure. 

\section{Conclusions and Outlook}
The present review shows that the PR-DNS approach is on an
equal footing with laboratory experimental measurements, and that
data-sets from both types of sources can be used in a complementary
fashion in order to improve our understanding of particulate flow
problems. For the future it can be expected that the application of
PR-DNS will further expand to encompass additional physical phenomena,
as well as growing system sizes which will then allow to cover the
large parameter space more extensively.
One obvious obstacle here is the sheer size of the data which has
made the process of extracting physical insight from raw simulation
output an arduous task.
\revision{Furthermore, the direct impact of PR-DNS results on reduced order
  models has thus far been only of relatively limited extent.}{%
  Furthermore, the research community has only relatively recently
  begun to utilize PR-DNS results for the purpose of model development
  \citep{tenneti:14}.} 
Applying modern methods from data science to the growing body of data
resulting from PR-DNS will provide a promising way forward 
(cf.\ related discussion in chapter~14). 
Another technique -- which has not been used extensively in this
context in the past -- is the method of ``numerical experiments''
\citep{moin:98}. The possibility to simulate the ``wrong
physics'' in PR-DNS in order to elucidate a specific mechanism is a
powerful technique indeed. Several authors have already employed this
method in the past (e.g.\ suppressing particle rotation in order to
find out its impact on the particle dynamics \citep{kajishima:04b};
varying the computational domain size in order to find threshold
length scales of the physical system, cf.\
\S~\ref{sec:PRR-wall-bounded-sediment-patterns}), 
and it is deemed potentially fruitful to design further numerical
experiments of this kind.
Finally, let us state yet another approach for reducing the
complexity of the system investigated with the aid of PR-DNS. Instead
of tackling fully developed turbulent flow, one can study particles
interacting with one of a growing set of known invariant solutions to
the Navier-Stokes equations~\citep{kawahara:12a,pestana:19}, some of
which have been shown to be statistically relevant to turbulence.
While the combination of PR-DNS and invariant solutions has not yet
demonstrated its full potential, it appears as a worthwhile avenue to
explore.  
%
%
%
%
%
%
%
%
%
%
%
%
%
%
%
%
%
%
%
%
%
%
%
%
%
\bibliographystyle{apa-good-doi}
\bibliography{TPFbook,MU3.bib,AGK3.bib,AC3.bib,JD3.bib}
\end{document}